\documentclass[a4paper,11pt]{article}

\usepackage{jheppub} 

\author[a]{Ibrahima Bah}
\author[a]{Ross Dempsey}
\author[a]{and Peter Weck}
\affiliation[a]{Department of Physics and Astronomy, Johns Hopkins University, \\
3400 North Charles Street, Baltimore, MD 21218, USA}

\emailAdd{iboubah@jhu.edu}
\emailAdd{srossd@jhu.edu}
\emailAdd{pweck1@jhu.edu}

\usepackage{graphicx}
\usepackage{dcolumn}
\usepackage{bm}
\usepackage[colorlinks=true,linkcolor=blue]{hyperref}
\usepackage{todonotes}
\usepackage{amsmath,amsthm,amsfonts}
\usepackage{mathtools}
\usepackage{tikz}
\usetikzlibrary{calc,decorations.markings,arrows}
\usepackage[title]{appendix}

\usepackage{titlesec}
 \usepackage{etoolbox, chngcntr}
\AtBeginEnvironment{appendices}{%
 \titleformat{\section}{\bfseries\Large}{\appendixname~\thesection:}{0.5em}{}%
 \titleformat{\subsection}{\bfseries\large}{\thesubsection}{0.5em}{}%
\counterwithin{equation}{section}
}

\renewcommand{\d}[2]{\frac{d #1}{d #2}}
\newcommand{\dd}[2]{\frac{d^2 #1}{d #2^2}}

\def\R{\mathbb{R}}

\DeclareMathOperator{\diag}{diag}
\let\Re\relax
\let\Im\relax
\DeclareMathOperator{\Re}{Re}
\DeclareMathOperator{\Im}{Im}

\begin{document}

\title{Kerr-Schild Double Copy and Complex Worldlines}

\date{\today}

\begin{abstract}
{
We use the classical double copy to identify a necessary condition for Maxwell theory sources to constitute single copies of Kerr-Schild solutions to Einstein's equations. In the case of four-dimensional Kerr-Schild spacetimes on Minkowski backgrounds, we extend this condition to a parameterization of the corresponding single copies. These are given by Li\'enard-Wiechert fields of charges on complex worldlines. This unifies the known instances of the Kerr-Schild double copy black holes on flat four-dimensional backgrounds into a single framework. Furthermore, we use the more generic condition identified to show why the black ring in five dimensions does not admit Kerr-Schild coordinates.
}
\end{abstract}

\maketitle
\newpage
\section{\label{sec:intro}Introduction}

The amplitudes program in quantum field theory has revealed new and unexpected connections between gauge theories and gravity. Through the double copy relationship, in which gravity amplitudes are closely tied to the squares of gauge theory amplitudes, it has become possible to compute gravity amplitudes which would otherwise be prohibitively complicated \citep{Bern2008NewAmplitudes,Bern2010GravityTheory,Bern2010PerturbativeTheory}. A natural question is whether a similar relationship holds between exact solutions, in which solutions to general relativity can be generated from solutions to a gauge theory. Indeed, using the Kerr-Schild transformation, a simple and precise relationship can be drawn between gauge fields and spacetime metrics 
\citep{
Kerr2009RepublicationEquations,
Monteiro2014BlackCopy,Luna2019TypeCopy,
Ilderton2018Screw-symmetricVortex,
Luna2015TheSpacetime,
Gonzalez2018TheSpacetimes,
Gonzalez2019TheDimensions,
Luna2016TheHoles}.
It is striking that many exact solutions to the Einstein equations can be presented as the double copies of gauge theory solutions, including all four-dimensional black hole spacetimes. 

Generically, nonlinear behavior in general relativity makes a boundary-value approach very difficult \citep{Choquet-Bruhat1969GlobalRelativity}. However, in electrodynamics this is the natural way to treat a system. As highlighted by the double copy prescription, Kerr-Schild spacetimes represent a sector of general relativity in which the metric can be derived as the solution to a boundary-value problem, just as in  electrodynamics. 

In this paper, we make use of the boundary-value character of Kerr-Schild geometries in identifying a simple necessary condition for Maxwell sources to admit a Kerr-Schild double copy. The corresponding field strength must admit a null geodesic eigenvector whose differential reproduces the field strength. That is,
\begin{equation} \label{eq:condition}
F^\nu_\mu A^\mu= \chi A^\nu, \qquad F_{\mu \nu}= \partial_\mu A_\nu-\partial_\nu A_\mu,
\end{equation}
for some scalar function $\chi$. In the case of four-dimensional Minkowski backgrounds, we extend this condition to provide a succinct parameterization of all four-dimensional black hole spacetimes. They are double copies of real slices of Maxwell fields sourced by point charges moving on complex worldlines, or complex Li\`enard-Wiechert fields \cite{Newman2004MaxwellCongruences}. Different restrictions on these complex worldlines give the double copies derived in \cite{Monteiro2014BlackCopy} and \cite{Luna2016TheHoles}.

In higher dimensions or on curved backgrounds, it is not clear whether an analogous classification of Kerr-Schild black hole spacetimes is possible. However, by interpreting \eqref{eq:condition} as a necessary condition on the trajectories of particles probing a putative single copy gauge current, this approach provides us a simple physical test for Kerr-Schild double copy structure. This test is demonstrated to exclude the five-dimensional black ring spacetime, furnishing a new proof that it does not admit a double copy presentation in terms of a Kerr-Schild metric.

In section \ref{sec:ks}, we review the properties of Kerr-Schild spacetimes and the classical double copy, and present a systematic formulation of the latter. We clarify how the current source in the gauge theory is related to the stress-energy tensor in gravity. In section \ref{sec:gauge_conditions}, we describe the requirements which must be satisfied by a gauge field single copy of a Kerr-Schild metric. Referring to previous work of Newman, in section \ref{sec:lienard_wiechert} we relate these single copies to real slices of complex Li\'enard-Wiechert fields \citep{Newman2004MaxwellCongruences}, and use this identification to systematically construct all four-dimensional black hole spacetimes on Minkowski backgrounds. In section \ref{sec:breaking}, we present a test for Kerr-Schild structure in any number of dimensions. For sources which pass this test, we can carry the technique further to derive their Kerr-Schild coordinates, effectively uplifting solutions of boundary value problems in electrodynamics to solutions in general relativity. This procedure is outlined in section \ref{sec:ks_coordinates}.


\section{\label{sec:ks}Kerr-Schild Metrics and the Classical Double Copy}

The classical double copy relates solutions in gauge theory to Kerr-Schild spacetimes in general relativity \cite{Monteiro2014BlackCopy}. In section \ref{sec:ks_metrics} we review  results concerning the Kerr-Schild geometries, first introduced in \cite{Kerr2009RepublicationEquations}. In section \ref{ks_dbl_copy}, we review the stationary double copy discovered in \cite{Monteiro2014BlackCopy}, and present a generalization. In section \ref{sec:gauge_conditions}, we derive conditions required for a Maxwell theory solution to be related to a gravity solution by a double copy, and present the problem of classifying Kerr-Schild spacetimes using their gauge theory counterparts.

\subsection{\label{sec:ks_metrics}Kerr-Schild Spacetimes}

A Kerr-Schild metric is obtained from a transformation on a fixed background metric $\overline{g}_{\mu\nu}$ of arbitrary dimension and curvature. Given a null vector field $k^\mu$ on this background, we can make the Kerr-Schild transformation
\begin{equation}\label{eq:ks}
    g_{\mu\nu} = \overline{g}_{\mu\nu} + \phi k_{\mu}k_{\nu},
\end{equation}
where $\phi$ is some scalar function. The spacetime with metric $g_{\mu\nu}$ is a Kerr-Schild spacetime and the null vector employed in the transformation is called the Kerr-Schild vector. For reasons to be explained shortly, we will consider only geodesic Kerr-Schild vectors.

Schematically we can see the double copy structure of these spacetimes by thinking of them as perturbations to background spacetimes in which the graviton is a tensor product of two copies of a null vector field. We will often refer to a Kerr-Schild spacetime as the full spacetime, in contrast to the background spacetime on which it is defined.

Contracting both sides of \eqref{eq:ks} by $k^\mu k^\nu$, we find that $g_{\mu\nu}k^\mu k^\nu = \overline{g}_{\mu\nu}k^\mu k^\nu$, so $k^\mu$ is also null with respect to $g_{\mu\nu}$. An important consequence is the truncation of the inverse metric to first order in the graviton,
\begin{equation}\label{eq:metric_inverse}
    g^{\mu\nu} = \overline{g}^{\mu\nu} - \phi k^\mu k^\nu.
\end{equation}
This truncation implies that $k^\mu$ is geodesic in the background spacetime if and only if it is geodesic in the full spacetime. It also allows us to make a simple statement of the condition for the Kerr-Schild vector to be geodesic. Kerr-Schild transformations change the Ricci tensor component $\overline{R}_{\mu \nu}k^\mu k^\nu$ by
\begin{equation}\label{eq:shift}
{R}_{\mu \nu}k^\mu k^\nu - \overline{R}_{\mu \nu}k^\mu k^\nu= \phi (k^\mu \overline{\nabla}_\mu k^\lambda)(k^\nu \overline{\nabla}_\nu k_\lambda).
\end{equation}
Note that we use bars throughout to refer to quantities defined with respect to the background spacetime. Therefore, if
\begin{equation} \label{eq:geodecity}
{R}_{\mu \nu}k^\mu k^\nu=\overline{R}_{\mu \nu}k^\mu k^\nu,
\end{equation}
then $k^\nu \overline{\nabla}_\nu k^\mu$ must be a null vector. Furthermore, $k_\mu k^\nu \overline{\nabla}_\nu k^\mu = 0$, so $k^\nu \overline{\nabla}_\nu k^\mu$ is then both null and orthogonal to $k^\mu$. This implies $k^\nu \overline{\nabla}_\nu k^\mu$ is proportional to $k^\mu$, i.e., that $k^\mu$ is geodesic. For example, if both the background and full spacetimes saturate the null energy condition, then $k^\mu$ is geodesic.

Kerr-Schild spacetimes with geodesic $k^\mu$ are most interesting, because of dramatic simplifications to their Ricci tensors. If we introduce a dimensionless perturbation parameter $\lambda$ into the Kerr-Schild transformation $g_{\mu \nu}=\overline{g}_{\mu \nu}+\lambda \phi k_\mu k_\nu$, we see that the truncation of the inverse metric $g^{\mu \nu}=\overline{g}^{\mu \nu}-\lambda \phi k^\mu k^\nu$ at first order implies that the Ricci tensor could be at most fourth order in $\lambda$. In fact, the Ricci tensor of a general Kerr-Schild spacetime with geodesic $k^\mu$ truncates at second order with lowered indices,
\begin{align}
R_{\alpha\beta} &= \overline{R}_{\alpha\beta} + \lambda R^{(1)}_{\alpha\beta} + \lambda^2 R^{(2)}_{\alpha\beta}, \\
    R^{(1)}_{\alpha\beta} &= \frac{1}{2}\overline{\nabla}_\sigma\left(\overline{\nabla}_\alpha(\phi k^\sigma k_\beta) + \overline{\nabla}_\beta (\phi k^\sigma k_\alpha) - \overline{\nabla}^\sigma(\phi k_\alpha k_\beta)\right),\\
    R^{(2)}_{\alpha\beta} &= \phi k_\alpha k^\sigma R^{(1)}_{\sigma\beta}.
\end{align}
Because of the form of the second-order term, we can raise one index and find
\begin{equation}
    {R^\alpha}_\beta ={\overline{R}^\alpha}_\beta - \lambda \left[ \phi k^\alpha k^\sigma \overline{R}_{\sigma\beta} - \overline{g}^{\alpha\sigma}R^{(1)}_{\sigma\beta} \right],
\end{equation}
a first-order truncation. Fixing the perturbation parameter to unity, the explicit mixed-index Ricci tensor for Kerr-Schild spacetimes (\ref{eq:ks}) with geodesic $k^\mu$ is
\begin{equation}\label{eq:ricci_truncation}
    {R^\alpha}_\beta = {\overline{R}^\alpha}_\beta  - \phi k^\alpha k^\sigma \overline{R}_{\sigma\beta} + \frac{1}{2}\overline{\nabla}_\sigma\left(\overline{\nabla}^\alpha(\phi k^\sigma k_\beta) + \overline{\nabla}_\beta(\phi k^\sigma k^\alpha) - \overline{\nabla}^\sigma(\phi k^\alpha k_\beta)\right),
\end{equation}
where ${\overline{R}^\alpha}_\beta = \overline{g}^{\alpha\gamma}\overline{R}_{\gamma\beta}$ and $\overline{\nabla}^\sigma = \overline{g}^{\sigma\lambda}\overline{\nabla}_\lambda$.

\subsection{\label{ks_dbl_copy}The Classical Double Copy}

It was recently discovered that stationary vacuum Kerr-Schild spacetimes are related to vacuum solutions of Maxwell's equations on the background spacetime \cite{Monteiro2014BlackCopy}. This follows directly from the form of the Ricci tensor given in \eqref{eq:ricci_truncation}. On a flat background the terms involving $\overline{R}_{\mu\nu}$ vanish, and so the vacuum Einstein equations give
\begin{equation}
	\overline{\nabla}_\sigma\left(\overline{\nabla}^\alpha(\phi k^\sigma k_\beta) + \overline{\nabla}_\beta(\phi k^\sigma k^\alpha) - \overline{\nabla}^\sigma(\phi k^\alpha k_\beta)\right) = 0.
\end{equation}
With some additional assumptions and gauge choices, we can find the Maxwell equations among these Einstein equations. We assume the spacetime is stationary and choose the stationary coordinates, in which $\overline{\nabla}_0(\phi k^\sigma k^\alpha) = 0$. Furthermore, we set $k_0 = 1$ by an appropriate choice of $\phi$, without changing the overall graviton. It follows that the Einstein equations with index $\beta = 0$ are
\begin{equation}
	\overline{\nabla}_\sigma\left(\overline{\nabla}^\alpha(\phi k^\sigma) -\overline{\nabla}^\sigma(\phi k^\alpha)\right) = 0.
\end{equation}
We see from this equation that a Kerr-Schild spacetime naturally defines a gauge field $A^\mu \equiv \phi k^\mu$. Indeed, if $\eta_{\mu\nu}+\phi k_\mu k_\nu$ is the metric of a stationary spacetime, the vacuum Einstein equations imply that $A^\mu$ solves the vacuum Maxwell equations. Note the Maxwell equations correspond only to ${R^\mu}_0 = 0$, and the other Einstein equations provide additional constraints on the Kerr-Schild graviton which are not related to the gauge field $A^\mu$.

We refer to the gauge field $A^\mu$ as the single copy of the metric $g_{\mu\nu}$, or more specifically, of the graviton $\phi k_\mu k_\nu$. An archetypal example of the single copy procedure is the relationship between the Schwarzschild metric and a Coulomb field. In Eddington-Finkelstein coordinates, the Schwarzschild metric is given by
\begin{equation}\label{eq:sch}
	ds^2 = \overline{ds}^2 + \frac{r_s}{r}\left(-dt+dr\right)^2,
\end{equation}
where $\overline{ds}^2$ is the line element of four-dimensional Minkowski space. The metric \eqref{eq:sch} is manifestly in Kerr-Schild form, with $\phi = \frac{r_s}{r}$ and $k^\mu = (\partial_t + \partial_r)^\mu$. The single copy gauge field is $A^\mu = \phi k^\mu$, and it satisfies
\begin{equation}
	\partial_\mu F^{\mu\nu} = j^\nu,
\end{equation}
where $F^{\mu\nu} = \partial^\mu A^\nu - \partial^\nu A^\mu$ and
\begin{equation}\label{eq:coulomb_current}
	j^\mu = -q \delta^{(3)}(x) (\partial_t)^\mu.
\end{equation}
Note that we have made the replacements $M\to q$, a charge, and $\kappa \to g$, the gauge coupling, when writing $A^\mu$, so that $r_s = \frac{\kappa M}{4\pi}$ becomes $\frac{gq}{4\pi}$.

In electrodynamics we think of a gauge field as the consequence of some configuration of current. This view is less applicable in general relativity, owing to its nonlinear behavior. However, the double copy relationship indicates that for Kerr-Schild spacetimes, it is instructive to think of a metric as the result of a source. Furthermore, we should think about how the gravity source is related to its corresponding gauge source. Indeed, the gauge current \eqref{eq:coulomb_current} is related to the source of the Schwarzschild metric,
\begin{equation}
	{T^\mu}_\nu = M\delta^{(3)}(x)(\partial_t)^\mu (dt)_\nu.
\end{equation}
If we follow the derivation of the double copy while keeping track of sources, a more general relationship becomes clear. From \eqref{eq:ricci_truncation}, we find that a stationary Kerr-Schild solution on flat background with $k_0 = 1$ satisfies
\begin{equation}
	{R^\mu}_0 = -\frac{M}{2q}\overline{\nabla}_\sigma F^{\sigma\mu}.
\end{equation}
Using the Einstein and Maxwell equations, this implies
\begin{equation}\label{eq:sources_dbl_copy_stationary}
	j^\mu = -\frac{2q}{M}\left({T^\mu}_0 - \frac{T}{D-2}{\delta^\mu}_0\right),
\end{equation}
where $T = {T^\mu}_\mu$. This result is also discussed in \cite{Gonzalez2019TheDimensions}.

\begin{figure}
    \centering
    \begin{tikzpicture}[
    decoration={markings,mark=at position 0.3 with {\arrow{triangle 60}},mark=at position 0.8 with {\arrow{triangle 60}}},scale=1.2
    ]
        \node at (0,0) (g) {$g_{\mu\nu}$};
        \node at (4,0) (T) {$T^\mu_\nu$};
        \node at (0,-4) (A) {$A^\mu$};
        \node at (4,-4) (j) {$j^\mu$};
        \draw[-latex] (g) to [bend left] node[above] {Einstein} (T);
        \draw[-latex,dashed] (T) to [bend left] node[below] {Einstein} (g);
        \draw[-latex] (A) to [bend left] node[above] {Maxwell} (j);
        \draw[-latex,dashed] (j) to [bend left] node[below] {Maxwell} (A);
        \draw[-latex] (g) -- node[rotate=90,above] {$A^\mu = \phi k^\mu$} (A);
        \draw[-latex] (T) -- node[rotate=90,below] {Eqs. \eqref{eq:sources_dbl_copy_stationary}, \eqref{eq:sources_dbl_copy}} (j);
        
    \end{tikzpicture}
    \caption{The relationships between the gravity solution $g_{\mu\nu}$, the gauge field $A^\mu$, the gravity source $T^\mu_\nu$, and the gauge source $j^\mu$. Dashed lines denote directions which require solving equations of motion.}
    \label{fig:double_copy_paths}
\end{figure}
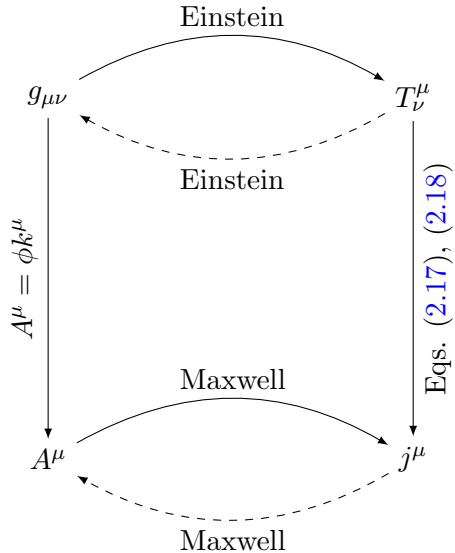

In the stationary case, this completes a web of relationships depicted in Figure \ref{fig:double_copy_paths}. Gravity, with the Einstein equations relating $g_{\mu\nu}$ to ${T^\mu}_\nu$, is shown as a layer above Maxwell theory, which relates $A^\mu$ to $j^\mu$. The classical double copy connects a Kerr-Schild metric $g_{\mu\nu}$ to $A^\mu$ by the prescription $A^\mu \equiv \phi k^\mu$. The sources are connected by \eqref{eq:sources_dbl_copy_stationary}, which we can use to construct the current $j^\mu$ from the stress-energy tensor ${T^\mu}_\nu$. Note that in both cases we are constructing elements of the gauge theory using elements of the gravity theory. Reconstructing a gravity solution from a gauge solution requires imposing additional constraints, which we explore in later sections.

The stationary Kerr-Schild spacetimes do not capture all known cases of the Kerr-Schild double copy. Double copy presentations have been given for accelerating black holes \cite{Luna2016TheHoles} and a pp-wave known as the vortex \cite{Ilderton2018Screw-symmetricVortex}, neither of which are stationary. Additionally, the double copy appears to work quite generally on a maximally symmetric background \cite{Gonzalez2018TheSpacetimes}. We will present a covariant version of the Kerr-Schild double copy, which applies in all these cases.

With stationary spacetimes, we treated the time coordinate as privileged and used the contraction $(\partial_t)^\nu {R^\mu}_\nu$ to obtain the Maxwell equations. We will generalize this here by replacing $(\partial_t)^\nu$ by a generic congruence $\ell^\nu$. Additionally, in order to treat non-flat backgrounds, we will keep track of the background Ricci tensor ${\overline{R}^\mu}_\nu$ and its source ${\overline{T}^\mu}_\nu$. The object of interest is then $\ell^\nu{(R-\overline{R})^\mu}_\nu$, and we find that the relevant gravity source is the stress-energy perturbation ${(T-\overline{T})^\mu}_\nu$.
We leave the  details of the derivation to Appendix \ref{sec:proofs}. The main point is that ${(R-\overline{R})^\mu}_\nu$ contains terms proportional to $\overline{\nabla}_\sigma F^{\mu\sigma}$, where $F^{\mu\sigma}$ is the field strength of $A^\mu \equiv \phi k^\mu$, and additional terms which vanish in the stationary case with $\ell^\mu = (\partial_t)^\mu$ but do not vanish generically. In order to construct the source of $A^\mu$, we need to keep track of these terms and add them to the current we get from contracting ${(T-\overline{T})^\mu}_\nu$ with $\ell^\nu$. The result is that $A^\mu$ solves the Maxwell equations on the background metric $\overline{g}_{\mu\nu}$ with source
\begin{equation}\label{eq:sources_dbl_copy}
    j^\mu = -\frac{2q}{M(\ell\cdot k)}\ell^\nu\left({(T-\overline{T})^\mu}_\nu - \frac{T-\overline{T}}{D-2}{\delta^\mu}_\nu\right) + \hat{\jmath}^\mu.
\end{equation}
The additional current $\hat{\jmath}^\mu$ is given by
\begin{align}\label{eq:jhat}
    \begin{split}
        \hat{\jmath} &= \ell^\nu {S^\mu}_\nu,\\
        {S^\mu}_\nu &= \frac{2q}{\kappa M(\ell\cdot k)}\Bigg(\phi k^\mu k^\sigma \overline{R}_{\sigma\nu} - \frac{\kappa M}{2qg}F^{\mu\sigma}\overline{\nabla}_\sigma k_\nu \\
        &\quad - \frac{1}{2}\overline{\nabla}_\sigma\left(\phi k^\sigma \left(\overline{\nabla}^\mu k_\nu + \overline{\nabla}_\nu k^\mu\right) + \phi k^\mu \left(\overline{\nabla}_\nu k^\sigma - \overline{\nabla}^\sigma k_\nu\right) + k^\mu k^\sigma \overline{\nabla}_\nu \phi\right)\Bigg).
    \end{split}
\end{align}

These equations summarize the most general possible Kerr-Schild double copy between Maxwell theory and pure Einstein gravity, or Einstein-Maxwell gravity if we include an electromagnetic tensor in $T_{\mu\nu}$. The Kerr-Schild double copy has also been studied for a more general Yang-Mills theory \cite{Monteiro2014BlackCopy}, for which the color degrees of freedom in the gauge field must be removed when forming the metric. Furthermore, several classical generalizations of the Kerr-Schild double copy prescription exist \cite{Luna2015TheSpacetime}, including a version based on curvatures rather than fields \cite{Luna2019TypeCopy}. We restrict attention to Kerr-Schild double copies of Maxwell theory, as formulated above. This instance of the double copy enjoys locality in the fields, unlike a BCJ double copy or even the classical double copy in \cite{Luna2019TypeCopy}, and makes for a particularly straightforward relationship between the gauge fields and the graviton.

We demonstrate a variety of examples using our prescription for the Kerr-Schild double copy of Maxwell theory in Appendix \ref{sec:applications}. In many cases things are made simplest by choosing $\ell^\mu$ to be a Killing field, such as in the stationary case. If $\ell^\mu$ is Killing with respect to both the background and full metrics, then we can rewrite the additional current as
\begin{equation}\label{eq:killing_mess}
    \hat{\jmath}^\mu = \frac{2q}{\kappa M(\ell\cdot k)}\left\lbrack \phi k^\mu (k_\sigma \overline{\nabla}^2 \ell^\sigma + \overline{R}_{\rho\sigma}k^\rho \ell^\sigma)- \overline{\nabla}_\sigma\left(\phi k^{[\sigma} v^{\mu]} \right)-\overline{\nabla}^{[\mu}(\phi k^{\sigma]})\overline{\nabla}_\sigma(\ell^\lambda k_\lambda)\right\rbrack.
\end{equation}

The double copy is most straightforward when $\hat{\jmath}^\mu = 0$. There are a number of conditions which must be satisfied to ensure a vanishing $\hat{\jmath}^\mu$. For example, if the background is flat, it is additionally required that $\ell^\mu$ be Killing in both the background and the Kerr-Schild spacetimes, and that $\ell^\mu$ and $\ell\cdot k$ both be background-covariantly constant. These assumptions hold in \cite{Monteiro2014BlackCopy}, where $\ell^\mu = (\partial_t)^\mu$ on a stationary spacetime and $\ell\cdot k = 1$.

\subsection{\label{sec:gauge_conditions}Single Copy Gauge Fields}

By construction, the Kerr-Schild double copy applies only to the restricted class of spacetimes with a Kerr-Schild form. However, it is less clear what class of gauge fields can be considered single copies of these Kerr-Schild spacetimes. Here we will develop necessary conditions for a gauge field to be realized as a single copy. These conditions will be used in Secs. \ref{sec:lienard_wiechert} and \ref{sec:ks_coordinates} to give two different perspectives on the landscape of Kerr-Schild geometries.

In order to admit a Kerr-Schild double copy, a gauge field must be geodesic and null:
\begin{align}
    A^\mu A_\mu &= 0 \qquad A^\mu \overline{\nabla}_\mu A^\nu = \chi A^\nu,
\end{align}
where $\chi$ is some scalar function. The norm of a Maxwell field is gauge-dependent, so satisfaction of the null condition requires us to fix the gauge appropriately.
Combining these two conditions, we find
\begin{align}\label{eq:gauge_field_eom}
\begin{split}
    \chi A^\nu = A^\mu \overline{\nabla}_\mu A^\nu &= A^\mu ({F_\mu}^\nu + \overline{\nabla}^\nu A_\mu) \\
    &= -{F^\nu}_\mu A^\mu.
\end{split}
\end{align}
Thus, for $A^\mu$ to be a null geodesic field, it must be an eigenvector of its field strength. Since the field strength is an antisymmetric tensor, all its eigenvectors are null. Thus we can simplify our double copy condition to a gauge-invariant statement: the field strength must have a geodesic eigenvector which serves as its own four-potential. 

In four dimensions this condition can be expressed elegantly in terms of the Maxwell field strength in the formalism of Newman and Penrose \cite{Newman-Penrose1962}. In particular, it implies that the field strength can be encoded by two independent components, rather than six. Appendix \ref{sec:newman_penrose} discusses this formulation of our necessary condition, and the corresponding simplifications to the Riemann tensor for spacetimes obtained via the double copy.

We can use this necessary condition to study the scope of Kerr-Schild spacetimes from a gauge theory perspective. In section \ref{sec:lienard_wiechert}, we identify additional conditions sufficient to generate all four-dimensional black hole spacetimes on Minkowski backgrounds via the Kerr-Schild double copy, including solutions describing accelerating black holes \cite{Luna2016TheHoles}. 
Later, in section \ref{sec:ks_coordinates}, we will return to the general case, and outline a procedure for generating the metric in Kerr-Schild form from the putative single copy of any given stress-energy distribution.

\section{\label{sec:lienard_wiechert}4-D Double Copies from Complex Li\'enard-Wiechert Fields}

The truncation of the Ricci tensor for Kerr-Schild spacetimes shows that for these special geometries, general relativity is essentially a linear theory. Many of the exact solutions in general relativity are of Kerr-Schild type, even if they were not originally derived in this form. A sample of these spacetimes is given in Appendix \ref{sec:applications}. It is natural to ask which sources give rise to Kerr-Schild geometries, and how we can systematically construct them.

Certainly if we choose any scalar function $\phi$, any null geodesic vector field $k^\mu$, and any background metric $\overline{g}_{\mu \nu}$ we can write down a Kerr-Schild metric. However, its source will not generically be of any particular physical interest. The Schwarzschild and Kerr solutions are special because they are vacuum Kerr-Schild spacetimes on Minkowski backgrounds. However, requiring a vacuum geometry is too restrictive; we are also interested in charged solutions, or in solutions containing gravitational radiation. That is, we will be interested in metrics sourced by the sum of a stress-energy tensor of the form $\Phi m^\mu m_\nu$, for null $m^\mu$, and an electromagnetic source with current vanishing almost everywhere.

To obtain Kerr-Schild metrics with stress-energy tensors of this form in four dimensions, the Kerr-Schild vector must be shear-free. To see this, let us work up from the vacuum case. As noted in the paper introducing the Kerr-Schild class of solutions, in the case of a vacuum Kerr-Schild spacetime the Goldberg-Sachs theorem implies that $k^\mu$ must be shear-free \cite{Kerr2009RepublicationEquations}. We can draw the same conclusion in a spacetime with source of form $\Phi m^\mu m_\nu$ with $m^\mu$ null, using an extended version of the Goldberg-Sachs theorem \cite{Nurowski2015AThree}. 

Furthermore, it turns out that charging a four-dimensional Kerr-Schild black hole solution corresponds to a rescaling of the function $\phi$ which leaves $k_\mu$ invariant. \footnote{In particular, charging a four-dimensional Kerr-Schild solution is a special case of the deformations $\phi\mapsto f\phi$ which leave the trace of the stress-energy tensor unchanged. From the Ricci scalar $R = \overline{\nabla}_\mu \overline{\nabla}_\nu\left(\phi k^\mu k^\nu\right)$, it follows that the shift $\phi\mapsto \phi f$ leaves $T$ invariant whenever
\begin{equation}
    \phi \ddot{f} + (2\dot{\phi}+2\phi\theta+\phi\chi)\dot{f} = 0,
\end{equation}
where dots denote $k^\mu \overline{\nabla}_\mu$, $\theta = \overline{\nabla}_\mu k^\mu$, and $k^\mu \overline{\nabla}_\mu k^\nu = \chi k^\nu$. Charged solutions are obtained by choosing an $f$ which solves this constraint.} In other words, the charged four-dimensional black holes are constructed with the same shear-free $k^\mu$ as their uncharged limiting cases.  Using the necessary condition identified in section \ref{sec:gauge_conditions}, we conclude that for a  spacetime sourced by some combination of a stress-energy tensor $\Phi m^\mu m_\nu$ and an electromagnetic source with current vanishing almost everywhere, the single copy field strength has a principal vector which is both geodesic and shear-free.


Returning to the question of systematically constructing the Kerr-Schild black hole sources, we could treat this as a problem in general relativity and directly formulate our conditions on the stress-energy tensor as a constraint on $\phi$ and $k^\mu$. For example, as just noted in four dimensions $k^\mu$ must be shear-free in addition to null and geodesic. However, we can gain more insight by using the double copy paradigm to map this to a problem in electromagnetism on flat spacetime. Indeed, in four dimensions the problem of finding suitable Maxwell fields is well-studied. In four dimensional Minkowski space, the Maxwell fields with a geodesic and shear-free principal vector can be constructed from the Li\'enard-Wiechert fields of charges on complex worldlines \citep{Newman2004MaxwellCongruences}. We will show that the Kerr-Schild double copies of Li\'enard-Wiechert fields reproduce the four-dimensional black hole spacetimes on Minkowski backgrounds, including accelerating black holes. A similar relationship has previously been identified between the Li\'enard-Wiechert field of a pair of point particles and the C-metric, a spacetime which can naturally be described as a pair of accelerating black holes \cite{Luna2019TypeCopy}.

In section \ref{sec:dirac_gauge} we relate the class of gauge fields described in section \ref{sec:gauge_conditions} to the real slices of Li\'enard-Wiechert potentials for a charge with a complex worldline, as described in \cite{Newman2004MaxwellCongruences}. On a Minkowski background, we can use Poincar\'e symmetry to break the possible worldlines into four essentially different types: real stationary, complex stationary, real accelerating, and complex accelerating. In section \ref{sec:flat_worldlines}, we show that the double copies of first three cases are Schwarzschild-type, Kerr-type, and bremsstrahlung-type geometries, respectively. We also consider what happens when we try to include a cosmological constant, or when we try to form a Kerr-Schild double copy of the most general complex accelerating worldlines.

\subsection{\label{sec:dirac_gauge}Li\'enard-Wiechert Single Copy Fields}

In the preceding sections, we showed that if a gauge solution is the single copy of a Kerr-Schild black hole spacetime, its field strength must have a geodesic and shear-free principal vector. Furthermore, this eigenvector must be parallel to a gauge field corresponding to the field strength. A classification of Maxwell fields on four-dimensional Minkowski space with these properties has been carried out in terms of complexified Li\'enard-Wiechert fields \cite{Newman2004MaxwellCongruences}. We briefly review this construction here.

The Li\'enard-Wiechert field is that of a particle with charge $q$ moving on a worldline $y^\mu(\tau)$, and can be constructed using the retarded position of the charge. At a given spacetime point $x^\mu$, let $y^\mu(\tau_\text{ret})$ be the unique spacetime position of the particle for which $x^\mu-y^\mu(\tau_\text{ret})$ is null, and let $\lambda^\mu(\tau_\text{ret}) = \left.\d{y^\mu}{\tau}\right|_{\tau_\text{ret}}$ be the worldline velocity at that point. We let $r$ denote the spatial distance to $x^\mu$ in the frame of the particle,
\begin{equation}
	r = \lambda^\mu(\tau_\text{ret})(x_\mu-y_\mu(\tau_\text{ret})).
\end{equation}
The Li\'enard-Wiechert gauge field is then a natural generalization of the Coulomb field,
\begin{equation}\label{eq:lw}
	A^\mu = \frac{gq}{4\pi r}\lambda^\mu(\tau_\text{ret}).
\end{equation}
The corresponding field strength has a null, geodesic, shear-free, and twist-free principal vector. In fact, any field strength with this property is a Li\'enard-Wiechert field.

The result of \cite{Newman2004MaxwellCongruences} is to generalize this construction to twisting congruences. If we take a complex extension of Minkowski space and consider a particle on a worldline $z^\mu(\tau) = x^\mu(\tau)+i y^\mu(\tau)$, \eqref{eq:lw} gives a complex gauge field which can be used to construct a complex field strength. Complex Maxwell fields bear a close relationship to their real counterparts. 
In terms of the complex field strength $\tilde F$, the corresponding real field strength is given by
\begin{equation}\label{eq:real_field_strength}
    F = \Re \tilde F - \Im \star \tilde F,
\end{equation}
where $\star$ is the Hodge dual taken with respect to complexified Minkowski space.
This allows us to construct a real solution to the vacuum Maxwell equations starting from a complex solution. If we apply this process to the complex Li\'enard-Wiechert field, the resulting real field has a null, geodesic, and shear-free principal vector. In fact, it is shown in \cite{Newman2004MaxwellCongruences} that any Maxwell field with such a principal vector is a real projection of a complex Li\'enard-Wiechert field.

\subsection{\label{sec:flat_worldlines}Construction of Double Copies}
In this section we construct the Kerr-Schild double copies of complex Li\'enard-Wiechert fields. We can compute the real principal null vectors of a Li\'enard-Wiechert field strength, and determine whether either of them are proportional to a gauge field generating that field strength. If we find such a gauge field $A^\mu$, then we can choose a splitting $A^\mu = \phi k^\mu$ and form a Kerr-Schild metric
\begin{equation}\label{eq:ks_gauge_field}
    g_{\mu\nu} = \overline{g}_{\mu\nu} + \phi k_\mu k_\nu.
\end{equation}
The choice of splitting can be thought of as a choice of $\phi$ in the graviton $\phi^{-1}A_\mu A_\nu$. 
For generic choices of the splitting, the source of this metric will bear little resemblance to the gauge source. Since the gauge current vanishes almost everywhere, we choose the splitting such that the metric is a solution to Einstein's equations with traceless stress-energy tensor. With this traceless condition are able to generate all four-dimensional black hole spacetimes on Minkowski background spacetimes as double copies of real slices of complex Li\'enard-Wiechert fields. 
 
Many of the examples of Appendix \ref{sec:applications} have different scalar functions $\phi$ but share the same null geodesic field $k^\mu$. In view of this, we define a ``generalized double copy'' of the gauge solution $A^\mu$ to be one which uses the vector $k^\mu$ obtained as described above, but a generic function $\phi$. If we require that $g_{\mu\nu}$ be a vacuum solution we recover the $\phi$ which satisfies $A^\mu = \frac{gq}{\kappa M}\phi k^\mu$; if we relax this requirement, we obtain other functions $\phi$ which are also of interest.

\subsubsection*{Stationary Worldlines}
We start with the real stationary worldline $\vec{x}$, which we can take to be at the origin. The Li\'enard-Wiechert gauge field is 
\begin{equation}
    A^\mu(\vec{x}) = \frac{gq}{4\pi r}\left((\partial_t)^\mu + (\partial_r)^\mu\right),
\end{equation}
where $r\equiv |\vec{x}|$. It is indeed an eigenvector of its own field strength tensor. It follows that $\phi k_\mu = \frac{r_s}{r}(-dt+dr)_\mu$, and so the Kerr-Schild metric is
\begin{equation}
    g_{\mu\nu} = \eta_{\mu\nu} + \frac{1}{\phi(r)} \frac{r_s^2}{r^2} (-dt+dr)_\mu (-dt+dr)_\nu.
\end{equation}
This metric has a stress-energy tensor with trace
\begin{equation}\label{stationary_trace}
    T = -\frac{\kappa^{-1}}{r^2}\dd{}{r}\left(r^2\phi(r)\right).
\end{equation}
In order to have a traceless source, which we could interpret as a Maxwell stress-energy tensor, we must have $\phi(r) = \frac{r_s}{r} - \frac{r_q^2}{r^2}$, where $r_s$ and $r_q$ are integration constants. This gives the Reissner-Nordstr\"om black hole, with the Schwarzschild black hole as a special case, as the double copy of a real stationary charge.

We can follow a similar but more involved analysis for the complex stationary worldline. In complex Minkowski coordinates $z^\mu = x^\mu + i y^\mu$, we take the position of the charge to be $-i\vec{a}$, where $\vec{\cdot}\ $ denotes a real 3-vector. The Li\'enard-Wiechert field is then simply the complex Coulomb field, with potential
\begin{equation}\label{eq:complex_potential}
    \varphi(\vec{x}) = \frac{gq}{4\pi\sqrt{(\vec{x}+i\vec{a})^2}}= \frac{gq}{4\pi}\frac{r-ia\cos\theta}{r^2+a^2\cos^2\theta},
\end{equation}
where we have used spheroidal coordinates for which $(\vec{x}+i\vec{a})^2 = (r+ia\cos\theta)^2$. 
Note that this complex field could also be obtained  from a transformation of the Coulomb potential taking $r \rightarrow r+i a \cos \theta$, the same replacement appearing in the Newman-Janis Trick transforming the Schwarzschild solution to Kerr \cite{Newman1965,erbin2017}. Indeed, as we are about to show, the double copy of the field in \eqref{eq:complex_potential} is the Kerr solution.

Using the potential in \eqref{eq:complex_potential}, we can construct the real field strength ${F^\mu}_\nu$, and compute its principal null vectors. They are $\left(1, \pm 1, 0, -\frac{a}{r^2+a^2}\right)$. Comparing with \eqref{eq:complex_potential}, we see that in order to have $A^0(\vec{x}) = \Re \varphi(\vec{x})$ we must set
\begin{equation}\label{complex_station_vec}
    A^\mu(\vec{x}) = \frac{gq}{4\pi}\frac{r}{r^2+a^2\cos^2\theta}\left(1, \pm 1, 0, -\frac{a}{r^2+a^2}\right).
\end{equation}
This is indeed the single copy gauge field of the Kerr metric, as discussed in Appendix \ref{sec:applications}.

Following \eqref{eq:ks_gauge_field}, we next find a function $\phi$ which makes the metric $\eta_{\mu\nu}+\phi^{-1}A_\mu A_\nu$ into a solution with traceless source. Starting from the generic graviton $\frac{1}{\phi(r,\theta)} A_\mu A_\nu$, we find that the trace of the source is given by
\begin{equation}
    T = -\frac{\kappa^{-1}}{r^2+a^2\cos^2\theta}\dd{}{r}\left(\left(r^2+a^2\cos^2\theta\right)\phi(r,\theta)\right),
\end{equation}
so we can obtain the traceless solution with
\begin{equation}
    \phi = \frac{r_s r - r_q^2}{r^2+a^2\cos^2\theta}
\end{equation}
where $r_s$ and $r_q$ are constants of integration. This gives the Kerr-Newman black hole with mass and charge set by $r_s$ and $r_q$.

\subsubsection*{Including a Cosmological Constant}
In the constructions presented so far, we have fixed the trace of the stress-energy tensor to zero. If instead we only require that it is some constant $\Lambda$, then for the real stationary family of double copies we can have solutions to (\ref{stationary_trace}) of the form
\begin{equation}\label{ads_sch}
    \phi(r) = \frac{\Lambda}{3}r^2 + \frac{r_s r -r_q^2}{r^2}.
\end{equation}
If we fix the entire stress-energy tensor by requiring ${T^\mu}_\nu = -\kappa \Lambda {\delta^\mu}_\nu$, then this $\phi$ is the unique solution to (\ref{stationary_trace}). With the Schwarzschild Kerr-Schild vector, (\ref{ads_sch}) is precisely the Kerr-Schild transformation from Minkowski space to the (A)dS-Schwarzschild black hole. Therefore (A)dS-Schwarzschild is also a generalized double copy of a real point charge.

However, for the case of the stationary complex worldline, there is no choice of $\phi$ which gives a source ${T^\mu}_\nu \propto {\delta^\mu}_\nu$. Simply demanding that the trace of the stress-energy tensor is constant requires a $\phi$ function of the form
\begin{equation}
\phi=\frac{\Lambda}{3} r^2 \left( \frac{r^2+6a^2 \cos^2 \theta}{r^2+a^2 \cos^2 \theta} \right)+\frac{r_s r-r_q^2}{r^2+a^2\cos^2\theta}.
\end{equation}
But with the Kerr-Schild vector (\ref{complex_station_vec}), this generates a spacetime with stress-energy tensor deviating from that of (AdS)-Kerr by terms of order two and higher in $1/r$. In other words, we cannot form the (AdS)-Kerr metric as a generalized double copy of a complex stationary charge in Minkowski space. This suggests that the appearance of the (A)dS-Schwarzschild metric in the real stationary family of Kerr-Schild metrics is accidental. It relies on the fact that (A)dS and Schwarzschild are required by their $SO(3)$ symmetry to share the same Kerr-Schild vector when taken as Kerr-Schild transforms of Minkowski space.

\subsubsection*{Real Accelerating Worldlines}
We now consider double copies of charges with real accelerating worldlines $y^\mu(\tau)$, with velocity $\lambda^\mu(\tau)$ and acceleration $\dot{\lambda}^\mu(\tau)$. The Li\'enard-Wiechert field is given by
\begin{equation}
    A^\mu = \left.\frac{gq}{4\pi r}\lambda^\mu\right|_\text{ret},
\end{equation}
where $r$ and the retarded time prescription are as in \ref{sec:bremsstrahlung}, such as \eqref{eq:lw}. Using \eqref{eq:br_identities}, we find that the field strength is given by
\begin{equation}\label{eq:br_field_strength}
    {F^\mu}_\nu = \frac{gq}{4\pi r}\left(k^\mu \alpha_\nu - \alpha^\mu k_\nu\right),
\end{equation}
where $\alpha^\mu = \dot{\lambda}^\mu - \frac{1+r(k\cdot\dot{\lambda})}{r}\lambda^\mu$.

By construction, the null geodesic $k^\mu$ tangent to light cones centered on the worldline is a principal null vector of the field strength. Here we note a major difference between accelerating and non-accelerating worldlines. If we write the gauge field as
\begin{equation}
    A^\mu = \frac{gq}{4\pi r}k^\mu + \frac{gq}{4\pi r}(\lambda^\mu-k^\mu),
\end{equation}
the second term is pure gauge if and only if $\dot{\lambda}^\mu = 0$. Indeed, denoting the terms composing the gauge field by $A^\mu \equiv A^\mu_0 + A^\mu_\text{rad}$, we have
\begin{align}\label{eq:magic_br_split}
\begin{split}
    {(F_0)^\mu}_\nu &= -\frac{gq}{4\pi r^2}\left(k^\mu \lambda_\nu - \lambda^\mu k_\nu\right),\\
    {(F_\text{rad})^\mu}_\nu &= \frac{gq}{4\pi r}\left(k^\mu \beta_\nu - \beta^\mu k_\nu\right)
\end{split}
\end{align}
with $\beta^\mu = \dot{\lambda}^\mu - (k\cdot\dot{\lambda})\lambda^\mu$. Since $F_0$ also has $k^\mu$ as a principal null vector, and $A_0^\mu \propto k^\mu$, we can form a double copy using $A_0^\mu$. This field has a source given by
\begin{equation}
    \partial_\mu F^{\mu\nu} = 0 \quad\implies\quad \partial_\mu F_0^{\mu\nu} = -\partial_\mu F^{\mu\nu}_\text{rad} = \frac{gq}{4\pi}\frac{2(k\cdot \dot{\lambda})}{r^2}k^\nu.
\end{equation}
We find the splitting $A_0^\mu = \phi k^\mu$ by using the trace-free condition $T=0$. In order to have $T = 0$, we must have
\begin{equation}
    R = \partial_\mu \partial_\nu \left(\phi^{-1} A_0^\mu A_0^\nu\right) = 0.
\end{equation}
Direct computation shows that $R\propto 2\phi'(r)-\phi(r)\phi''(r)$, which vanishes for $\phi(r) = \frac{c_1}{r+c_2}$, where $c_1$ and $c_2$ are constants. Therefore, the double copy is given by
\begin{equation}\label{charged_accel_bh}
    g_{\mu\nu} = \eta_{\mu\nu} + \frac{r_s r-r_q^2}{r^2}k_\mu k_\nu,
\end{equation}
where $r_s$ and $r_q$ are related to the constants of integration. If we set the constant $r_q=0$ this reproduces the accelerating black hole of \cite{Luna2016TheHoles}, discussed in Appendix \ref{sec:bremsstrahlung}. The metric (\ref{charged_accel_bh}) appears to be a charged generalization of the accelerating black hole. To confirm this hypothesis, we compute the full stress-energy tensor,
\begin{equation}
    {T^\mu}_\nu = \frac{3r_s (k\cdot\dot{\lambda})}{\kappa r^2}k^\mu k_\nu - \frac{r_q^2}{\kappa r^4}\left\lbrack 2(1-rk\cdot\dot{\lambda})k^\mu k_\nu + {\delta^\mu}_\nu - 2(\lambda^\mu k_\nu +k^\mu \lambda_\nu)\right\rbrack.
\end{equation}
If $\dot{\lambda} = 0$, this reduces to $\frac{r_q^2}{r^4}\diag(-1, -1, 1, 1)$, which is indeed the source for the Reissner-Nordstr\"om metric.

\subsubsection*{\label{sec:complex_br}Complex Accelerating Worldlines}

The stationary and real accelerating worldlines considered above are all special cases of complex accelerating worldlines, the most general source for complex Li\'enard-Wiechert fields. Since complex displacement of the worldline corresponds to rotation of the black hole, we expect the complex accelerating worldline to give a rotating black hole with nonzero linear and angular acceleration.

This would in principle be a very general class of spacetimes. If we take the acceleration to be real and parallel to the complex shift, then we would have a Kerr black hole which accelerates along its spin axis. This resembles the source of the rotating C-metric, which is the most general of the Plebanski-Demianski metrics with vanishing NUT charge \cite{GRIFFITHS2006ASOLUTIONS}. These metrics are of Petrov type D, but generically they are not Kerr-Schild.

This suggests that the fortuitous splitting in \eqref{eq:magic_br_split}, in which the radiative piece of the gauge field can be split off to leave behind a field strength which admits a Kerr-Schild double copy, may not occur in the complex case. This would prevent us from viewing the real slice of the Li\'enard-Wiechert field of a complex accelerating worldline as the single copy of a black hole spacetime. Nonetheless, the relation between complex accelerating charges and the Plebanski-Demianski solution is worthy of further study. A generalization of the Kerr-Schild double copy, which holds for any type D spacetime, has been described in terms of curvatures rather than fields \cite{Luna2019TypeCopy}. Using this double copy of curvatures, the authors noted that the C-metric can be presented as the double copy of a pair of accelerating point charges. It is conceivable that generic Plebanski-Demianski metrics could be understood as curvature double copies of complex Li\'enard-Wiechert fields.


\section{\label{sec:breaking} Constraints on Double Copy Sources}
In section \ref{sec:gauge_conditions}, we found that a gauge vector can only be used to form a Kerr-Schild double copy in Einstein gravity if it is both geodesic and an eigenvector of its field strength. This holds independently of the additional constraints to the stress-energy tensor we employed in section \ref{sec:lienard_wiechert}.

We now demonstrate that this condition gives us a physical picture explaining why the five-dimensional rotating black ring source does not admit a Kerr-Schild metric. We anticipate that analogous tests can be applied to other topologically non-trivial black hole spacetimes in higher dimensions to prove that, like the black ring, they do not admit the Kerr-Schild double copy presentation enjoyed by four-dimensional black holes.

\subsection{The Worldline Scattering Test}
We can put more physical intuition behind the double copy condition \eqref{eq:gauge_field_eom} by treating 
\begin{align} \label{eq:gauge_field_condition}
    {F^\nu}_\mu A^\mu = -\chi A^\nu.
\end{align}
as an expression for the Lorentz force on a particle with worldline velocity $u^\mu \equiv A^\mu$ \citep{Caser2001ElectrodynamicsEquivalence}. We fix the charge to mass ratio of the particle at -1, and then take the limit $m\to 0$ to make sense of nullity condition $u_\mu u^\mu = 0$. If the gauge field is geodesic, then $u^\mu$ is everywhere an eigenvector of ${F^\mu}_\nu$, so we use an eigenvector as the initial velocity at some point far from the source. The particle then evolves according to
\begin{align}\label{eq:gauge_trajectory}
    \d{x^\mu}{\tau} &= u^\mu, \qquad \d{u^\mu}{\tau} = {F^\mu}_\nu(x(\tau))u^\nu.
\end{align}
The gauge field is geodesic only if the trajectory of the particle is a geodesic curve. In particular, on a Minkowski background, straight line trajectories indicate that the gauge field is geodesic.

This test compares the trajectories of massless on-shell particles in the pure gravitational background $\overline{g}_{\mu\nu}$ to particles probing an additional gauge field $A^\mu$. Curiously we find that for single copy gauge fields, there is a family of null curves which are on-shell in both of these backgrounds, namely the curves which are everywhere tangent to an eigenvector of the gauge field strength. This is an unexpected relationship, and it places strict constraints on the gauge fields which can be double copied.

\begin{figure}
    \centering
    \includegraphics[width=\linewidth]{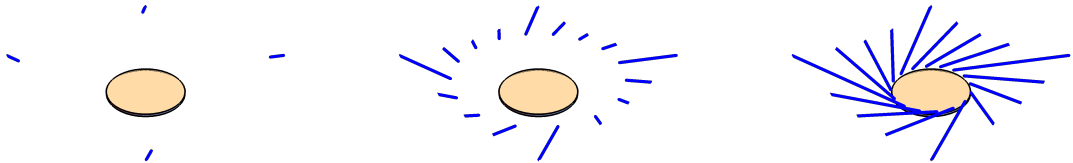} 
    \caption{Numerical solutions to the equations \eqref{eq:gauge_trajectory} for the single copy of the Kerr geometry at three time steps, showing that the trajectories are straight lines.}
 \label{fig:kerr_trajectories}
\end{figure}

As an example, we take another look at the Kerr geometry. Using \eqref{eq:sources_dbl_copy} to identify the single copy $j^\mu$ of the Kerr stress-energy distribution, we can integrate $-\overline{\nabla}^2 A^\mu = j^\mu$ to find the gauge field in Lorenz gauge, $\tilde A^\mu$. The result, derived in Appendix \ref{sec:kerr_ks}, is that 
\begin{equation}\label{eq:Atilde_Kerr}
\tilde A^\mu= \frac{gq}{4\pi}\frac{r}{r^2+a^2\cos^2\theta}\left(1,0,0,-\frac{a}{r^2+a^2} \right)
\end{equation}
Since the field strength is gauge invariant, we can compute it and its eigenvectors using $\tilde{A}^\mu$. In this case, the two real eigenvectors of ${F^\mu}_\nu$ are
\begin{equation}
    v^\mu_\pm = \frac{gq}{4\pi}\frac{r}{r^2+a^2\cos^2\theta}\left(1, \pm 1, 0, -\frac{a}{r^2+a^2}\right).
\end{equation}
When we take the positive sign, this is the gauge field single copy of the Kerr metric found in \ref{sec:kerr}. The negative sign gives an equivalent double copy. Both choices are gauge-equivalent to $\tilde A^\mu$.

To apply the worldline scattering test, we compute the field strength of \eqref{eq:Atilde_Kerr} and use it to write the equations of motion in \eqref{eq:gauge_trajectory}. To set the initial conditions, we fix a spacetime point and set the initial velocity to $v^\mu_-$ at that point. Integrating \eqref{eq:gauge_trajectory} gives the trajectories shown in Figure \ref{fig:kerr_trajectories}. As we expect, the particles follow straight line paths, corresponding to geodesic curves on the Minkowski background. 

\begin{figure}
  \centering
    \includegraphics[width=\linewidth]{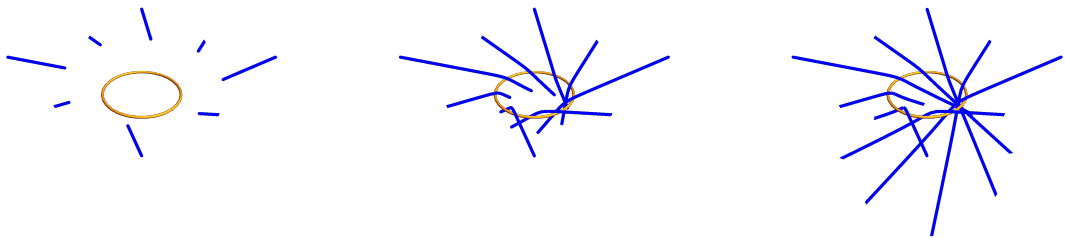}
    \caption{Numerical solutions to the equations \eqref{eq:gauge_trajectory} for the single copy of the black ring geometry at three time steps. Scattering in the vicinity of the ring shows that the gauge current does not admit a null geodesic gauge field.}
   \label{fig:ring_trajectories}
\end{figure}

\subsection{\label{sec:black_ring}Application to the Black Ring}
Thanks to uniqueness theorems, in four dimensions the Kerr-Newman family exhausts the possibilities for asymptotically flat black hole solutions. In section \ref{sec:lienard_wiechert} these solutions were all generated as double copies of a class of gauge fields. However, in higher dimensions, more diverse singular objects are permitted and expected to exist \citep{Emparan2008BlackDimensions}. In five dimensions, a black hole with $S^1\times S^2$ horizon topology has been identified \citep{Emparan2006BlackRings}. This spacetime, known as the black ring, has no Kerr-Schild form on a flat background and thus does not admit a Kerr-Schild double copy description in Minkowski space.

The lack of a Kerr-Schild form for the black ring can be proved using the algebraic classification of spacetimes by their Weyl-aligned null directions, or WANDs, as noted in \citep{Ett_2010}. A Kerr-Schild transformation on a maximally symmetric background produces an algebraically special spacetime, with $k^\mu$ a multiple WAND \cite{Malek2011}. The black ring is of more general Petrov type I.

We offer a physical explanation of this fact using the scattering test developed in the previous subsection. We take massless charged particles and give them an initial worldline velocity aligned with an eigenvector of the field strength for the single copy of the black ring suggested by (\ref{eq:sources_dbl_copy}). If the solutions to \eqref{eq:gauge_trajectory} are not straight lines, then there is no hope of finding a gauge transformation which makes the gauge field null and geodesic.

Figure \ref{fig:ring_trajectories} shows that the particle trajectories are scattered in the vicinity of the ring. This establishes that, after making a putative single copy of the black ring source, there is no way to form a Kerr-Schild double copy and complete the cycle in Figure \ref{fig:double_copy_paths}.

Indeed, from the perspective of the worldline scattering test, it is not surprising that the black ring should fail. In the Kerr geometry, a disk is removed from the spacetime, and so particle trajectories end on this disk. This allows for a twisting geodesic field, in which trajectories above and below the disk twist in the same direction without having to bend and connect. For the black ring, however, only an infinitesimal ring is removed from the spacetime, and so our test particles pass through the plane of the ring. Smoothly connecting the twisting geodesic fields from above and below this plane would not be possible, and so we should expect the test particles to scatter.

\subsection{\label{sec:ks_coordinates}Integrating Gravitational Sources}

When a source passes the worldline scattering test, we can carry the process further and determine a Kerr-Schild solution to the Einstein equations for this source. This represents a boundary value approach to gravity, in which we map a gravitational source to a gauge current, integrate the gauge current in electrodynamics, and then lift the resulting gauge field to generate a metric.

We will exhibit this process for the Schwarzschild metric. In the usual coordinates \eqref{eq:sch}, the Schwarzschild metric is sourced by
\begin{equation}
    T^\mu_\nu = M\delta^{(3)}(x)(\partial_t)^\mu (dt)_\nu. 
\end{equation}
In order to form the single copy of this source, we employ a Killing vector $\ell$. Since the source is static, we can use the constant timelike Killing vector. Additionally, we need to know what $\hat{\jmath}$ will be. We will follow the prescription $\ell\cdot k = -1$, so that the constancy of $\ell$ and $\ell\cdot k$ imply $\hat{\jmath} = 0$. We then have
\begin{equation}
    j^\mu = -q\delta^{(3)}(x)(\partial_t)^\mu. 
\end{equation}
Solving the Maxwell equations in Lorenz gauge gives
\begin{equation}
    \tilde A^\mu = \int d^3 r'\, \frac{-gj^\mu(r')}{4\pi|r-r'|} = \frac{gq}{4\pi r}(\partial_t)^\mu.
\end{equation}
Forming a field strength from this gauge field, we find that its real principal vectors are $\partial_t \pm \partial_r$. Via \eqref{eq:gauge_field_condition}, this suggests we use the gauge field
\begin{equation}
	A^\mu = \frac{gq}{4\pi r}\left(\partial_t \pm \partial_r\right)^\mu,
\end{equation}
and indeed we find that $A^\mu$ and $\tilde A^\mu$ differ by pure gauge. In order to form a double copy, we split the gauge field into $\frac{gq}{\kappa M}\phi k^\mu$ such that $\ell\cdot k = -1$. This splitting implies that the Kerr-Schild transformation is given by
\begin{align}\label{eq:schwarzschild_transformation}
    \phi &= \frac{r_s}{r}, \qquad k^\mu = (1, \pm 1, 0, 0).
\end{align}
The sign ambiguity derives from a choice between retarded and advanced potentials, as can be seen in section \ref{sec:bremsstrahlung}. We will take the positive sign, corresponding to a retarded potential. 

This process alone is not enough to guarantee \emph{a priori} that \eqref{eq:schwarzschild_transformation} gives the Schwarzschild geometry; we would still need to check that the resulting metric solves the full Einstein equations. However, we can easily recognize that \eqref{eq:schwarzschild_transformation} is the correct Kerr-Schild transformation for the Schwarzschild metric, and so we forego these steps here.

A similar approach applied to the source of the Kerr metric yields the Kerr-Schild form of this metric. Owing to the singular behavior of the source near its outer ring, the integration of the gauge current is considerably more complicated than for Schwarzschild. We present the details in Appendix \ref{sec:kerr_ks}. 

These two examples are suggestive of a more general technique for integrating gravitational sources by means of the classical double copy. The main difficulty is in forming the gauge current. Since the Schwarzschild and Kerr solutions are stationary, we could use the timelike Killing vector and normalize $k^\mu$ to ensure that $\hat{\jmath} = 0$ in \eqref{eq:sources_dbl_copy}. More generally, we need some way of computing $\hat{\jmath}$ or proving that it vanishes before we can proceed with the method outlined here.


\section{\label{sec:discussion}Discussion and Conclusion}

Our classification of Kerr-Schild solutions on a Minkowski background in section \ref{sec:lienard_wiechert} applies only in four dimensions. While this result does provide a better understanding of the Kerr-Schild sector of general relativity, and its relationship to gauge theory, the solutions we encounter are already well-understood. Black hole solutions of general relativity in four dimensions can be neatly classified using uniqueness theorems. These theorems do not extend to five and higher dimensions, and so it would be of great interest to develop our classification technique in this context.

The key to extending our procedure to higher dimensions would be generalizing Newman's theorem to classify the Maxwell fields with geodesic and shear-free principal vectors in higher dimensions. One of these Maxwell fields is the single copy of the Myers-Perry black hole, a generalization of the Kerr black hole to arbitrary dimensions which admits a Kerr-Schild form \cite{Monteiro2014BlackCopy}. However, the use of a complex shift to generate twisting Maxwell fields does not easily extend to generating the Myers-Perry single copy. In four dimensions, when shifting by $i\vec{a}$, the vector $\vec{a}$ was dual to the angular momentum 2-form. Such duality does not hold in other dimensions, and so a different formalism would be required to generate the twisting Maxwell fields in $D\neq 4$.

There is some evidence that, in lieu of complex numbers, some other division ring might suffice to produce the Myers-Perry solution. The so-called Newman-Janis trick, which generates the Kerr solution from the Schwarzschild solution via complex coordinate redefinitions, can be extended to the five-dimensional Myers-Perry solution using quaternions \cite{Mirzaiyan2017GeneratingQuaternions}. The full quaternion algebra is not necessary for this construction; we only need to extend $\R$ by a set of scalars which square to -1 and anticommute amongst themselves (in this case, $\{i, j\}$). This possibility is worthy of further exploration, and may aid in generating some subset of the Kerr-Schild black hole solutions in higher dimensions.

We can also consider Kerr-Schild solutions on curved backgrounds. Many instances of the classical double copy have been described on maximally symmetric background \cite{Gonzalez2018TheSpacetimes}. In four dimensions, we might expect a complexification of the (A)dS manifold to yield a method for generating the (A)dS-Kerr solution via a complex shift of the (A)dS-Schwarzschild solution. The primary difficulty is determining what we mean by a complex shift in this case. In Minkowski space, the affine structure allows us to shift a worldline by $i\vec{a}$, but there is no such structure in (A)dS space.

Recent work suggests that this may be the consequence of a more fundamental difficulty. Using the classical double copy, the complex shift which generates the Kerr solution has been explained in terms of minimal-coupling amplitudes in momentum space \cite{Arkani-Hamed2019KerrParticles}. The lack of affine structure in (A)dS space, which makes the idea of a complex shift ambiguous, also impedes the use of momentum space formalism. It is thus unclear whether the (A)dS-Kerr solution can be thought of as a complex shift of the (A)dS-Schwarzschild solution.

These difficulties aside, our conclusions are as follows. Using the classical double copy to explore the Kerr-Schild sector of general relativity, we show that Kerr-Schild solutions on four-dimensional flat backgrounds can be constructed as double copies of complex Li\'enard-Wiechert fields \cite{Newman2004MaxwellCongruences}. This provides a clear organizing principle behind the Kerr-Schild forms of the Schwarzschild, Kerr, and bremsstrahlung geometries \citep{Monteiro2014BlackCopy,Luna2016TheHoles}, as well as their charged analogs. It sheds greater light on the relationship between Li\'enard-Wiechert fields and general relativity, first studied in \cite{Lind1974} and \cite{Newman1974}, by connecting to the broader double copy paradigm.

Further work is needed to clarify the extent to which similar methods might be applicable in higher dimensions or on maximally symmetric spaces. Nonetheless, we do present a necessary condition on the sources of Kerr-Schild metrics, the worldline scattering test, which holds on any background in any spacetime dimension. This test uses the classical double copy to map a stress-energy tensor to a gauge current. If a massless charged particle probing this gauge current is scattered off of a geodesic trajectory, then the stress-energy tensor cannot be associated with a Kerr-Schild geometry. This test is stringent enough to show that the black ring in five dimensions does not admit a Kerr-Schild metric.

It is conceivable that further constraints or generating techniques for Kerr-Schild solutions can be derived using the classical double copy. Thinking of Kerr-Schild metrics as double copies of gauge fields provides a new and illuminating way of studying geometries which have long been known to be especially simple, owing to linearization of the Einstein equations, yet which harbor a rich class of exact black hole solutions. Furthermore, it may be possible to extend the blueprint of Fig. \ref{fig:double_copy_paths} to the more general Weyl double copy, thus providing insight into all spacetimes of Petrov type D \cite{Luna2019TypeCopy}. Finally, it is of great interest to connect these developments in the classical double copy more robustly to the BCJ double copy, such as in \cite{Luna2016TheHoles}. We leave all these questions to future work.


\section*{Acknowledgments}

We thank 
Nima Arkani-Hamed for initial discussions that lead to this project.  We also thank Jared Kaplan, David Kaplan, Harold Erbin, Cindy Keeler, Tucker Manton, and Federico Bonetti 
for interesting conversations and correspondence. The work of IB, RD, and PW is supported in part by NSF grant PHY-1820784.


\begin{appendices}


\section{Single Copy Gauge Currents \label{sec:proofs}}

Here we provide a more detailed derivation of \eqref{eq:sources_dbl_copy}. The Einstein equations for $g_{\mu\nu}$ and $\overline{g}_{\mu\nu}$ can be expressed as
\begin{align}\label{eq:efe}
    R^\mu_\nu &= \kappa\left(T^\mu_\nu - \frac{T}{D-2}\delta^\mu_\nu\right), \qquad  \overline{R}^\mu_\nu = \kappa\left(\overline{T}^\mu_\nu - \frac{\overline{T}}{D-2}\delta^\mu_\nu\right).
\end{align}
Additionally, since $g_{\mu\nu} = \overline{g}_{\mu\nu}+\phi k_{\mu}k_{\nu}$, \eqref{eq:ricci_truncation} says that
\begin{equation}\label{eq:start}
    R^\mu_\nu - \overline{R}^\mu_\nu = -\phi k^\mu k^\sigma \overline{R}_{\sigma\nu} + \frac{1}{2}\overline{\nabla}_\sigma \left(\overline{\nabla}_\nu(\phi k^\sigma k^\mu) + \overline{\nabla}^\mu(\phi k^\sigma k_\nu) - \overline{\nabla}^\sigma(\phi k^\mu k_\nu)\right).
\end{equation}
The left hand side can be rewritten in terms of ${T^\mu}_\nu$ using \eqref{eq:efe}. On the right hand side, we pull out the terms
\begin{equation}
    \frac{1}{2}k_\nu \overline{\nabla}_\sigma \left(\overline{\nabla}^\mu(\phi k^\sigma)-\overline{\nabla}^\sigma(\phi k^\mu)\right) = \frac{\kappa M}{2q}j^\mu k_\nu.
\end{equation}
The remaining terms on the right hand side form the tensor 
\begin{align}
\begin{split}
    &\phi k^\mu k^\sigma \overline{R}_{\sigma\nu} - \frac{\kappa M}{2qg}F^{\mu\sigma}\overline{\nabla}_\sigma k_\nu \\
        & - \frac{1}{2}\overline{\nabla}_\sigma\left(\phi k^\sigma \left(\overline{\nabla}^\mu k_\nu + \overline{\nabla}_\nu k^\mu\right) + \phi k^\mu \left(\overline{\nabla}_\nu k^\sigma - \overline{\nabla}^\sigma k_\nu\right) + k^\mu k^\sigma \overline{\nabla}_\nu \phi\right).
\end{split}
\end{align}
Thus, upon contracting \eqref{eq:start} with a congruence $\ell^\nu$, we find
\begin{equation}
    \ell^\nu\left({(T-\overline{T})^\mu}_\nu - \frac{T-\overline{T}}{D-2}{\delta^\mu}_\nu\right) = \frac{M(\ell\cdot k)}{2q}(j^\mu + \ell^\nu {S^\mu}_\nu),
\end{equation}
using the definition of $S^\mu_\nu$ from \eqref{eq:jhat}. Rearranging, \eqref{eq:sources_dbl_copy} follows.

If we have a vector $\ell^\nu$ which is Killing with respect to both $g_{\mu\nu}$ and $\overline{g}_{\mu\nu}$, then we can simplify further. The Killing equations imply $\ell^\nu \overline{\nabla}_\nu(\phi k^\sigma k^\mu) = 2\phi k^{(\mu} \dot{\ell}^{\sigma)}$. Applying this to \eqref{eq:start}, we find
\begin{align}
    \kappa \ell^\nu\left({T^\mu}_\nu - {\overline{T}^\mu}_\nu \right. &+ \left. \frac{T-\overline{T}}{D-2}{\delta^\mu}_\nu\right)  \nonumber \\ &= \frac{1}{2}\overline{\nabla}_\sigma \left(2\phi k^{(\mu}\dot{\ell}^{\sigma)} + \ell^\nu \overline{\nabla}^\mu(\phi k^\sigma k_\nu) - \ell^\nu \overline{\nabla}_\sigma(\phi k^\mu k_\nu)\right) - \phi k^\mu (\overline{R}_{\sigma\nu}k^\sigma \ell^\nu ) \nonumber  \\
    &- \frac{1}{2}(\overline{\nabla}_\sigma \ell^\nu)\left(\overline{\nabla}_\nu(\phi k^\sigma k^\mu) + \overline{\nabla}^\mu(\phi k^\sigma k_\nu) - \overline{\nabla}^\sigma(\phi k^\mu k_\nu)\right).
\end{align}
In the first term, if we move $\ell^\mu$ inside the derivatives, we obtain something in the form of a field strength tensor. If we define the gauge field
\begin{equation}
    A^\mu \equiv -\frac{gq}{\kappa M}\phi k^\mu,
\end{equation}
then the relevant term is $\frac{\kappa M}{2qg}\overline{\nabla}_\sigma F^{\mu\sigma}$, where $F^{\mu\sigma} = \overline{\nabla}^\mu A^\sigma - \overline{\nabla}^\sigma A^\mu$. In total we have
\begin{align}
     \kappa \ell^\nu\left({T^\mu}_\nu - {\overline{T}^\mu}_\nu \right. &+ \left. \frac{T-\overline{T}}{D-2}{\delta^\mu}_\nu\right) \nonumber \\ &=  \frac{\kappa M}{2qg}\left((\ell^\lambda k_\lambda)\overline{\nabla}_\sigma F^{\mu\sigma} + F^{\mu\sigma}\overline{\nabla}_\sigma(\ell^\lambda k_\lambda)\right) \nonumber \\
    &+\frac{1}{2}\overline{\nabla}_\sigma\left(\phi k^\sigma (\ell^\lambda \overline{\nabla}^\mu k_\lambda - k_\lambda \overline{\nabla}^\mu \ell^\lambda) - \phi k^\mu (\ell^\lambda \overline{\nabla}^\sigma k_\lambda - k_\lambda \overline{\nabla}^\sigma \ell^\lambda)\right) \nonumber \\
    &-\phi k^\mu (k_\sigma \overline{\nabla}^2 \ell^\sigma + \overline{R}_{\sigma\nu}k^\sigma \ell^\nu)
\end{align}
where we have simplified using the background Killing equation for $\ell^\mu$. Using the Maxwell equations $\overline{\nabla}_\sigma F^{\mu\sigma} = gj^\mu$, we obtain the desired result.

\section{\label{sec:newman_penrose} Single Copy Field Strengths}
It can be useful to study the double copy at the level of the field strengths, in addition to the fields. This can be done most easily in the formalism of Newman and Penrose \cite{Newman-Penrose1962}. In this formalism, tensors are projected onto a complete vector basis, consisting of a null tetrad of two real null vectors $\{ l^\mu, n^\mu \}$ and two complex conjugate null vectors $\{m^\mu, \overline{m}^\mu \}$ satisfying the orthogonality conditions
\begin{equation}
m_\mu n^\mu=m_\mu l^\mu =0, \qquad l_\mu n^\mu=m_\mu \overline{m}^\mu=1.
\end{equation} 
This formalism is particularly well-suited for studying the asymptotics of the Riemann tensor, with $\{ l^\mu, n^\mu \}$ asymptotically aligned with ingoing and outgoing null rays. These asymptotics are encoded by five complex Newman-Penrose invariants, related to tetrad components of the Riemann tensor  \cite{Debney1969,Goldberg2009}
\begin{align}
C^{(5)} &= 2 R_{\mu \nu \rho \sigma}l^\mu m^\nu l^\rho m^\sigma \\
C^{(4)} &=  R_{\mu \nu \rho \sigma}n^\mu l^\nu l^\rho m^\sigma+R_{\mu \nu \rho \sigma} \bar{m}^\mu m^\nu l^\rho m^\sigma \\
C^{(3)} &= 2R_{\mu \nu \rho \sigma}l^\mu m^\nu n^\rho \bar{m}^\sigma+\frac{R}{6} \\
C^{(2)} &=R_{\mu \nu \rho \sigma}n^\mu l^\nu n^\rho \bar{m}^\sigma+R_{\mu \nu \rho \sigma} \bar{m}^\mu m^\nu n^\rho \bar{m}^\sigma \\
C^{(1)} &= 2 R_{\mu \nu \rho \sigma}n^\mu \bar{m}^\nu n^\rho \bar{m}^\sigma.
\end{align}
Which of these five complex quantities vanish determines the degeneracy of the Weyl tensor, and thus the Petrov type of the spacetime \cite{Petrov2000}. The Newman-Penrose formalism can be used to study the Maxwell field strength in an analogous way. The six real components of $F_{\mu \nu}$ are encoded by three complex components, 
\begin{align}
\Phi_0 &= F_{\mu \nu} l^\mu m^\nu \\
\Phi_1&=\frac{1}{2} F_{\mu \nu} (l^\mu n^\nu+\overline{m}^\mu m^\nu)\\
\Phi_2 &= F_{\mu \nu}\overline{m}^\mu n^\nu.
\end{align}
Recall that the Kerr-Schild double copy required the field strength to have a geodesic eigenvector which serves as its own four-potential. In the Newman-Penrose formalism, we can choose the real tetrad vector $l^\mu$ to align with this geodesic eigenvector. As noted in \cite{Newman2004MaxwellCongruences}, this eigenvector assumption requires $\Phi_0 = -\chi l^\mu m_\mu =0$. The assumption that $l^\mu$ is geodesic independently implies that $\Phi_2 = 0$. Therefore a Maxwell field which admits a Kerr-Schild double copy can be represented by a single complex field strength component, $\Phi_1$.

The Kerr-Schild double copy relates a gauge field to a metric, and so there should also be relationship between the field strength and the spacetime curvature. Indeed, this perspective is taken from the outset in the Weyl double copy \cite{Luna2019TypeCopy}, which reduces to the Kerr-Schild double copy in some cases. From our perspective, we can see this relationship in terms of the Newman-Penrose invariants of the Weyl tensor.

In choosing a null tetrad for a spacetime obtained from the Kerr-Schild double copy, we can align one of the real null tetrad vectors $l^\mu$ with the null and geodesic Kerr-Schild vector $k^\mu$. When such a spacetime is defined on a Minkowski background, or any background in which the Weyl Tensor has at least one repeated eigenvector, two Newman-Penrose invariants vanish automatically \cite{Bilge1983}. In our conventions \cite{Debney1969,Goldberg2009}, $C^{(4)}=C^{(5)}=0$. This means that Kerr-Schild spacetimes on algebraically special backgrounds are themselves Petrov type II or more special. Many interesting Kerr-Schild spacetimes are more special, often of Petrov type D. Here type D implies $C^{(1)}=C^{(2)}=C^{(4)}=C^{(5)}=0$. In other words, only one Newman-Penrose invariant, $C^{(3)}$, is nonzero for all Kerr-Schild metrics on Minkowski backgrounds. It is related to the non-vanishing component of the Maxwell field strength for the corresponding single copy by
\begin{equation}
C^{(3)} =2Z \Phi_1+\frac{R}{6},
\end{equation}
where $R$ is the Ricci scalar and $Z$ is the complex parameter \cite{Debney1969,Goldberg2009}
\begin{equation}
Z=\theta+i \omega
\end{equation}
constructed from the real rates of expansion $\theta$ and twist $\omega$ of the geodesic null congruence defined by the Kerr-Schild vector. As a simple example, the single non-vanishing tetrad component of the Riemann tensor for the Schwarschild metric is 
\begin{equation}
C^{(3)} =-\frac{M}{r^3}
\end{equation}
while the single non-vanishing tetrad component of the corresponding single copy field strength is
\begin{equation}
\Phi_1= -\frac{q}{2r^2},
\end{equation}
where we have made the single copy replacement $M \rightarrow q$. The proportionality between the two components is fixed exactly by the expansion parameter
\begin{equation}
Z=r^{-1}.
\end{equation}
This correspondence between the Schwarzschild and Coulomb field strengths was observed long before the discovery of the classical double copy, such as in \cite{Newman1965}, and was extended to relate Kerr-Newman metrics to analogous Maxwell solutions \cite{Lind1974}.
The double copy places it within a broader context.

\section{\label{sec:applications}Kerr-Schild Examples}

Following the introduction of the classical double copy \citep{Monteiro2014BlackCopy}, other works have demonstrated instances of the relationship between Kerr-Schild spacetimes and gauge theory solutions \citep{Luna2016TheHoles,Luna2015TheSpacetime,Ilderton2018Screw-symmetricVortex,Chen2007Kerr-SchildMetrics,Gonzalez2018TheSpacetimes,Gonzalez2019TheDimensions}. In most instances, the gravity and gauge sources are obtained via their respective equations of motion and compared. Here we collect a representative sample of Kerr-Schild spacetimes and examine their gauge theory duals via \eqref{eq:sources_dbl_copy}. In each case we determine the gauge theory source via two methods: (i) substituting the gravity source directly into (\ref{eq:sources_dbl_copy}); and (ii) substituting the gauge field $A^\mu=-\frac{gq}{\kappa M}\phi k^\mu $ into Maxwell's equations and computing its source. Figure \ref{fig:double_copy_paths} shows the relationships between the solutions and sources in the gravity and gauge theories.

    \subsection{\label{sec:rn}Reissner-Nordstr\"om}
    
If we replace the Schwarzschild Kerr-Schild function $\phi = \frac{r_s}{r}$ with $\frac{r_s}{r}-\frac{r_q^2}{r^2}$, then Einstein's equations give a stress-energy tensor
\begin{equation}
    {T^\mu}_\nu = \frac{r_q^2}{\kappa r^4}\diag(-1,-1,1,1).
\end{equation}
The electromagnetic stress-energy tensor for a point charge $q$ at the origin is
\begin{equation}
    {(T_\text{EM})^\mu}_\nu = \frac{q^2}{32\pi^2 r^4}\diag(-1,-1,1,1).
\end{equation}
Therefore, we set $r_q^2\equiv \frac{\kappa q^2}{32\pi^2}$ and the Reissner-Nordstr\"om metric
\begin{equation}
    g_{\mu\nu} = \eta_{\mu\nu}+\phi k_\mu k_\nu,\qquad \phi = \frac{r_s}{r}-\frac{r_q^2}{r^2},\qquad k_\mu = (-1, 1, 0,0)
\end{equation}
is interpreted as the solution of the Einstein-Maxwell equations for a mass $M$ with charge $q$ sitting at the origin.

We know from the Schwarzschild case that the single copy of the $\frac{r_s}{r}$ piece is a point charge at the origin. By linearity, we are free to take $r_s = 0$ and consider only the charge piece here. We will use $A^\mu = \frac{2g}{\kappa} \phi k^\mu$ for the single copy gauge field, since the factor of $\frac{q}{M}$ is not needed in the present case. The gauge source obtained from Maxwell's equations is
\begin{equation}
    j^\mu = -\left(\frac{q}{4\pi r^2}\right)^2 (\partial_t)^\mu.
\end{equation}
We could also obtain this via equation (\ref{eq:sources_dbl_copy}), using the constant timelike Killing vector $(\partial_t)^\mu$. Just as in the previous example, appropriate conditions are satisfied to guarantee $\hat{\jmath}^\mu = 0$. Since the electromagnetic stress-energy tensor is traceless and the background is flat, we obtain
\begin{equation}
    j^\mu = 2\ell^\nu {T^\mu}_\nu = -\left(\frac{q}{4\pi r^2}\right)^2 (\partial_t)^\mu.
\end{equation}

    \subsection{\label{sec:kerr}Kerr}
    
In the usual Boyer-Lindquist coordinates, the Kerr metric for a rotating black hole is
\begin{align}
\begin{split}\label{eq:kerr_bl}
    ds^2 =\,& -\left(1-\frac{r_s r}{\Sigma}\right)\,dt^2 + \frac{\Sigma}{\Delta}\,dr^2 + \Sigma\,d\theta^2 \\
    &\left(r^2 + a^2 + \frac{r_s r a^2}{\Sigma}\sin^2\theta\right)\sin^2\theta\,d\varphi^2 - \frac{2r_s r a \sin^2\theta}{\Sigma}\,dt\,d\varphi,
\end{split}
\end{align}
where
\begin{align}
    \Sigma &= r^2 + a^2 \cos^2 \theta, \qquad \Delta = r^2 - r_s r + a^2.
\end{align}
The coordinates $(r, \theta, \varphi)$ are related to Cartesian coordinates by
\begin{align}\label{eq:spheroidal}
    x &= \sqrt{r^2+a^2}\sin\theta\cos\varphi, \qquad y = \sqrt{r^2+a^2}\sin\theta\sin\varphi, \qquad z = r\cos\theta.
\end{align}

The metric can be cast into Kerr-Schild form on a flat background in spheroidal coordinates with background metric
\begin{equation}
    \overline{g}_{\mu\nu}\,dx^\mu\,dx^\nu = -dt^2 + \frac{r^2+a^2\cos^2\theta}{r^2+a^2}\,dr^2 + (r^2+a^2\cos^2\theta)\,d\theta^2 + (r^2+a^2)\sin^2\theta\,d\varphi^2.
\end{equation}
\footnote{Note that this is the $r_s\to 0$ limit of Boyer-Lindquist coordinates.} The Kerr-Schild transformation is given by
\begin{align}\label{eq:kerr_ks}
    \phi &= \frac{r_s r}{r^2 + a^2 \cos^2\theta}, \qquad k^\mu = \left(1, 1, 0, -\frac{a}{r^2+a^2}\right).
\end{align}
This reduces to the Schwarzschild Kerr-Schild transformation as $a\to 0$. To form the single copy, we again use the constant timelike Killing field. Substituting $A^\mu = \frac{gq}{\kappa M}\phi k^\mu$ into the Maxwell equations gives the current source \citep{Monteiro2014BlackCopy}
\begin{equation}\label{eq:kerr_current}
    j^\mu = -\delta(r)\frac{q}{4\pi a^2}\sec^3\theta\left(2, 0,0,\frac{2}{a}\right).
\end{equation}

Again, we can derive this current source using (\ref{eq:sources_dbl_copy}). In Boyer-Lindquist coordinates, the source of the Kerr metric is given by \cite{Israel1970SourceMetric}
\begin{equation}
	T^\mu_\nu = \sigma\left(w^\mu w_\nu + \zeta^\mu \zeta_\nu\right),
\end{equation}
where $\sigma = \frac{M}{4\pi a^2\cos\theta}\delta(r)$, and the vectors appearing are
\begin{align}
	w_\mu &= \tan\theta(-1, 0, 0, a), \qquad \zeta_\mu = (0, 0, a\cos\theta, 0).
\end{align}
It follows that
\begin{equation}
	{T^\mu}_0 = -\sigma\left(\tan^2\theta, 0, 0, \frac{\sec^2\theta}{a}\right).
\end{equation}
Additionally, $w^\mu w_\mu = \zeta^\mu \zeta_\mu = 1$, so $T = 2\sigma$. The background is flat and $\ell^\mu$ and $\ell^\mu k_\mu$ are both constant, so it follows that
\begin{align}
    j^\mu &= -\frac{2q}{M}\left({T^\mu}_0 - \frac{T}{2}{\delta^\mu}_0\right) \\
    &= -\delta(r)\frac{q}{4\pi a^2}\sec^3\theta\left(2, 0, 0, \frac{2}{a}\right),
\end{align}
in agreement with \eqref{eq:kerr_current}.

    \subsection{\label{sec:pp-wave}Vortex}
    
A pp-wave spacetime is any spacetime which admits a covariantly constant null vector $k^\mu$. This vector field defines a null coordinate, $u \equiv k^\mu x_\mu$; together with an orthogonal null coordinate $v$, and two orthogonal spacelike coordinates $x$ and $y$, a pp-wave spacetime can be written in Brinkmann coordinates as
\begin{equation}
    ds^2 = 2\,du\,dv + dx^2 + dy^2 + \phi(u, x, y)\,du^2.
\end{equation}
The vacuum Einstein equations require $\partial_x^2 \phi + \partial_y^2 \phi = 0$. In Brinkmann coordinates, the pp-wave spacetime is manifestly in Kerr-Schild form, with $k_\mu dx^\mu = du$.

If we choose
\begin{equation}
    \phi = \phi_0\left(\cos(\omega u)(x^2-y^2) + 2\sin(\omega u)xy\right),
\end{equation}
then in addition to five Killing vectors which generally belong to pp-wave spacetimes, a sixth ``screw symmetry'' \cite{Ilderton2018Screw-symmetricVortex} appears, given by
\begin{equation*}
    \ell^\mu = (\partial_u)^\mu + \frac{\omega}{2}\left(x(\partial y)^\mu - y(\partial x)^\mu\right).
\end{equation*}
This screw symmetry can be used to construct a single copy. Although $\ell^\mu$ is not constant, it is still straightforward to show that $\hat{\jmath}^\mu = 0$. Since ${T^\mu}_\nu = 0$, equation (\ref{eq:sources_dbl_copy}) then tells us that the single copy is a solution to the vacuum Maxwell equations. Thus, there is no use in keeping track of constants. The gauge field is
\begin{equation}\label{eq:vortex_gauge_field}
    A^\mu \propto \phi k^\mu = \phi_0\left(\cos(\omega u)(x^2-y^2) + 2\sin(\omega u)xy\right)(\partial_v)^\mu,
\end{equation}
and indeed $\partial_\mu (\partial^{(\mu}A^{\nu)}) = 0$.
    
    \subsection{\label{sec:ads}(A)dS}

The maximally symmetric (anti)-de Sitter spacetime is given by the metric
\begin{equation}\label{eq:ads_standard}
    ds^2 = -\left(1-\frac{\Lambda r^2}{3}\right)\,dt^2 + \left(1-\frac{\Lambda r^2}{3}\right)^{-1}\,dr^2 + r^2\,d\Omega^2.
\end{equation}
If we make the coordinate transformation
\begin{equation}
    \overline{t} = -t - r + \sqrt{\frac{3}{\Lambda}}\tanh^{-1}\left(\sqrt{\frac{\Lambda}{3}}r\right),
\end{equation}
then this can be written in Kerr-Schild form as \citep{Chen2007Kerr-SchildMetrics}
\begin{equation}
    ds^2 = -d\overline{t}^2 + dr^2 + r^2\,d\Omega^2 + \frac{\kappa \rho_M}{3} r^2 (-d\overline{t}+dr)^2,
\end{equation}
where $\Lambda \equiv \kappa \rho_M$. This spacetime is sourced by $T_{\mu\nu} = -\rho_M g_{\mu\nu}$. Thus, rather than multiplying by a ratio of charges $\frac{q}{M}$ to form the single copy, it makes sense here to multiply by a ratio of charge densities $\frac{\rho_q}{\rho_M}$. We obtain
\begin{equation}
    A^\mu = \frac{g\rho_q}{\kappa \rho_M}\phi k^\mu (\ell^\nu k_\nu) = \frac{g\rho_q}{3}r^2(\partial_{\overline{t}}+ \partial_r)^\mu
\end{equation}
Substituting this into the Maxwell equations gives a gauge current
\begin{equation}\label{eq:ds_current}
    j^\mu = 2\rho_q (\partial_{\overline{t}})^\mu,
\end{equation}
corresponding to a uniform space-filling charge density. Indeed, according to equation (\ref{eq:sources_dbl_copy}) using the Killing vector $\ell^\mu = (\partial_t)^\mu$, the gauge current should be
\begin{align}
    j^\mu &= \frac{2\rho_q}{\rho_M(\ell^\lambda k_\lambda)}\ell^\nu\left({T^\mu}_\nu - \frac{T}{D-2}{\delta^\mu}_\nu\right) \\
    &= -2\rho_q\ell^\nu\left({\delta^\mu}_\nu - \frac{D}{D-2}{\delta^\mu}_\nu \right) \\
    &= 2\rho_q\ell^\mu,
\end{align}
which coincides with \eqref{eq:ds_current}.
    
    \subsection{\label{sec:ads-schwarzschild}(A)dS-Schwarzschild}
    
In Sect. \ref{sec:ks} and \ref{sec:ads}, we found Kerr-Schild transformations on a flat background which give the Schwarzschild and (anti)-de Sitter geometries, respectively. Both of these transformations have the same Kerr-Schild vector, $k_\mu\,dx^\mu = -dt + dr$. It is therefore simple to make both transformations at once by adding their scalar functions $\phi$. The result is the (A)dS-Schwarzschild metric in Kerr-Schild coordinates,
\begin{equation}\label{eq:ads_schwarzschild_ks}
    ds^2 = -dt^2 + dr^2 + r^2\,d\Omega^2 + \left(\frac{r_s}{r} + \frac{\kappa \rho_m}{3}r^2\right)(-dt+dr)^2.
\end{equation}
Making a single copy with the timelike Killing vector, we find a gauge field
\begin{equation}\label{eq:ads_schwarzschild_flat_bg}
    A^\mu = \left(\frac{gq}{4\pi r}+ \frac{g\rho_q}{3}r^2\right)(\partial_t + \partial_r)^\mu,
\end{equation}
with source
\begin{equation}\label{eq:ads_schwarzschild_gauge_source_flat_bg}
    j^\mu = (-q\delta^{(3)}(x)+2\rho_q)(\partial_t)^\mu,
\end{equation}
a superposition of the Schwarzschild and (A)dS gauge sources.

We can alternatively think of the (A)dS-Schwarzschild solution as a Kerr-Schild transformation on a curved background. In this case it is more illuminating to take the (A)dS space in the coordinates \eqref{eq:ads_standard}. In these coordinates, the Schwarzschild Kerr-Schild term takes the form
\begin{equation}
    ds^2 = ds_\text{(A)dS}^2 + \frac{r_s}{r}\left(dt + \frac{dr}{1-\frac{\Lambda}{3}r^2}\right)^2.
\end{equation}
The single copy gauge field is given by
\begin{equation}\label{eq:ads_schwarzschild_ads_bg}
    A^\mu = \frac{g q}{\kappa M}\phi k^\mu = \frac{gq}{4\pi r}\left(\frac{\partial_t}{1-\frac{\Lambda r^2}{3}} +\partial_r\right)^\mu.
\end{equation}
This gauge field should satisfy Maxwell's equations in an (A)dS background. Indeed, if we compute the field strength tensor $F^{\mu\nu} = 2\overline{\nabla}^{[\mu}A^{\nu]}$, the only nonzero component is
\begin{equation}
    F^{tr} = \frac{gq}{4\pi r^2},
\end{equation}
which is the radial electric field we should expect from a point charge at the origin. The divergence is
\begin{equation}\label{eq:ads_schwarzschild_gauge_source_ads_bg}
    j^\mu = \overline{\nabla}_\nu F^{\mu\nu} = -q\delta^{(3)}(x)(\partial_t)^\mu,
\end{equation}
the point source piece of \eqref{eq:ads_schwarzschild_gauge_source_flat_bg}. 

We can use equation (\ref{eq:sources_dbl_copy}) with the Killing vector $\ell^\mu = (\partial_t)^\mu$ to determine this gauge source, but the background is not Ricci flat, and $\ell^\mu$ is not covariantly constant, so every term in \eqref{eq:killing_mess} could in principle contribute. The only nonzero components of $\overline{\nabla}_\mu \ell_\nu$ are
\begin{equation}
    \overline{\nabla}_r \ell_t = -\overline{\nabla}_t \ell_r = \frac{\Lambda r}{3}.
\end{equation}
It follows that
\begin{equation}
    v^\mu = -\frac{\Lambda r}{3}k^\mu,
\end{equation}
so $k^{[\mu}v^{\nu]} = 0$. Additionally, $\ell^\lambda k_\lambda = -1$, so the last term vanishes. Finally, it is straightforward to show $k_\nu \overline{\nabla}^2 \ell^\nu = -\Lambda (\ell \cdot k)$ and $(\overline{R}_{\rho\sigma}k^\rho \ell^\sigma) = \Lambda (\ell \cdot k)$, so the first term vanishes as well. In total, $\hat{\jmath} = 0$.

Both eqs. \eqref{eq:ads_schwarzschild_gauge_source_flat_bg} and \eqref{eq:ads_schwarzschild_gauge_source_ads_bg} are consistent with (\ref{eq:sources_dbl_copy}). In both cases, the full stress-energy tensor is
\begin{equation}\label{eq:ads_schwarzschild_stress}
    {T^\mu}_\nu = -\rho_M \delta^\mu_\nu + M\delta^{(3)}(x)(\partial_t)^\mu(dt)_\nu. 
\end{equation}
On a flat background, ${\overline{T}^\mu}_\nu = 0$ and (\ref{eq:sources_dbl_copy}) gives eq. \eqref{eq:ads_schwarzschild_gauge_source_flat_bg}. On the (A)dS background, ${\overline{T}^\mu}_\nu = -\rho_M \delta^\mu_\nu$, so we only single copy the second term of \eqref{eq:ads_schwarzschild_stress}, and (\ref{eq:sources_dbl_copy}) gives \eqref{eq:ads_schwarzschild_gauge_source_ads_bg}.
    
    \subsection{\label{sec:ads-kerr}(A)dS-Kerr}
    
Following \cite{Gonzalez2018TheSpacetimes}, we can write the metric for a rotating black hole in (A)dS space using the background spacetime
\begin{gather}
    \overline{g}_{\mu\nu}\,dx^\mu\,dx^\nu = -\frac{\Delta}{\Omega}\left(1-\frac{\Lambda r^2}{3}\right)\,dt^2 + \frac{r^2 + a^2\cos^2\theta}{r^2+a^2} \frac{dr^2}{1-\frac{\Lambda r^2}{3}} \\+ \frac{r^2+a^2\cos^2\theta}{\Delta}\,d\theta^2 + \frac{(r^2+a^2)\sin^2\theta}{\Omega}\,d\varphi^2,
\end{gather}
where
\begin{align}
    \Delta &= 1+\frac{\Lambda}{3}a^2\cos^2\theta, \qquad \Omega = 1+\frac{\Lambda}{3}a^2.
\end{align}
The Kerr-Schild transformation is given by
\begin{align}
    \phi &= \frac{r_s r}{r^2+a^2\cos^2\theta}, \qquad k_\mu dx^\mu = -\frac{\Delta}{\Omega}\,dt + \frac{r^2+a^2\cos^2\theta}{r^2+a^2}\frac{dr}{1-\frac{\Lambda r^2}{3}} - \frac{a \sin^2\theta}{\Omega}\,d\varphi.
\end{align}

The single copy gauge field is $A^\mu = \frac{gq}{\kappa M}\phi k^\mu$. The corresponding field strength is
\begin{equation}\label{eq:ads_kerr_fields}
    F^{\mu\nu} = \frac{gq(r^2-a^2\cos^2\theta)}{4\pi (r^2+a^2\cos^2\theta)^3}
    \begin{pmatrix}
    0 & r^2+a^2 & -\frac{a^2 r \sin 2\theta}{r^2-a^2\cos^2\theta} & 0 \\
    -(r^2+a^2) & 0 & 0 & a\left(1-\frac{\Lambda}{3}r^2\right) \\
    \frac{a^2 r \sin 2\theta}{r^2-a^2\cos^2\theta} & 0 & 0 & -\frac{2ar\Delta\cot\theta}{r^2-a^2\cos^2\theta} \\
    0 & -a\left(1-\frac{\Lambda}{3}r^2\right) & \frac{2ar\Delta\cot\theta}{r^2-a^2\cos^2\theta} & 0
    \end{pmatrix}.
\end{equation}
Note that only the radial electric field survives in the limit $a\to 0$. This field strength satisfies the Maxwell equations in (A)dS background with source \cite{Gonzalez2018TheSpacetimes}
\begin{equation}\label{eq:ads_kerr_source}
    j^\mu = \frac{q}{4\pi a^2}\sec^3\theta \delta(r) \left(2,0,0,-\frac{2}{a}\right).
\end{equation}

We can also determine the gauge source using (\ref{eq:sources_dbl_copy}). We begin by showing that $\hat{\jmath} = 0$. Since the spacetime is maximally symmetric, $\overline{R}_{\mu\nu}k^\mu \ell^\nu = \Lambda (k\cdot \ell)$. It can also be shown that $k_\mu \overline{\nabla}^2 \ell^\mu = -\Lambda (k\cdot \ell)$, so the first term of $\hat{\jmath}$ vanishes. The vector $v^\mu$ is
\begin{equation}
    v^\mu = -\frac{r\Lambda}{3}\left(\frac{1}{1-\frac{\Lambda}{3}r^2}, \frac{r^2+a^2}{r^2+a^2\cos^2\theta}\frac{\Delta}{\Omega}, 0, 0\right),
\end{equation}
which gives
\begin{equation}
    \overline{\nabla}_\sigma\left(k^{[\sigma}v^{\mu]}\right) = \frac{r\Lambda}{3} \frac{\kappa m a^3}{2\pi}\frac{\cos^2\theta}{(r^2+a^2\cos^2\theta)^3}\frac{1}{\Omega}\left(a\sin^2\theta,0,0,-\Delta\right).
\end{equation}
We also have $\overline{\nabla}_\sigma(k\cdot \ell) = \frac{a^2\Lambda}{3\Omega}\sin 2\theta\left(0,0,1,0\right)$, which together with \eqref{eq:ads_kerr_fields} gives
\begin{equation}
    (\overline{\nabla}^{[\mu}(\phi k^{\sigma]}))\overline{\nabla}_\sigma(k\cdot \ell) = - \overline{\nabla}_\sigma\left(k^{[\sigma}v^{\mu]}\right) ,
\end{equation}
so these two terms in $\hat{\jmath}$ cancel each other, leaving $\hat{\jmath} = 0$.

Thus, to find the gauge source, we only need to evaluate ${T^\mu}_\nu$. In \cite{Israel1970SourceMetric}, the extrinsic curvature of the disk at $r = 0$ is used to evaluate the stress-energy tensor of the Kerr metric. Using the same method for the (A)dS-Kerr metric, we find that it is sourced by
\begin{equation}
    {T^\mu}_\nu = -\delta(r) \frac{M}{4\pi a^2}\sec\theta \left(w^\mu w_\nu + \zeta^\mu \zeta_\nu \right) - \frac{\Lambda}{\kappa} \delta^\mu_\nu,
\end{equation}
where
\begin{align}
    w_\mu &= \frac{\sqrt{\Delta}\tan\theta}{\Omega}\left(-1,0,0,a\right), \qquad \zeta_\mu = \frac{1}{\sqrt{\Delta}}\left(0,0,a \cos\theta, 0\right).
\end{align}
The second term is simply ${\overline{T}^\mu}_\nu$, so we subtract it off. We then have
\begin{equation}
    \ell^\nu\left({T^\mu}_\nu-{\overline{T}^\mu}_\nu\right) = \delta(r) \frac{M}{4\pi a^2}\sec\theta  (\ell\cdot w)w^\mu = \delta(r)\frac{M}{4\pi a^2}\sec^3\theta\frac{1}{\Omega}\left(-\sin^2\theta, 0, 0, \frac{\Delta}{a}\right).
\end{equation}
Both $w$ and $\zeta$ are unit vectors, so $\frac{T-\overline{T}}{D-2} = \frac{M}{4\pi a^2}\sec\theta\delta(r)$.  Thus, in total, we have
\begin{align}
    j^\mu &= -\frac{2q}{M(\ell\cdot k)}\frac{M}{4\pi a^2}\sec^3\theta\frac{1}{\Omega}\delta(r)\left(\sin^2\theta+\Omega\cos^2\theta, 0, 0, \frac{\Delta}{a}\right) \\
    &= \frac{1}{\Delta/\Omega}\frac{q}{4\pi a^2}\sec^3\theta\frac{1}{\Omega}\delta(r)\left(2\Delta, 0, 0, 2\frac{\Delta}{a}\right) \\
    &= \frac{q}{4\pi a^2}\sec^3\theta \delta(r)\left(2, 0, 0, \frac{2}{a}\right),
\end{align}
in agreement with \eqref{eq:ads_kerr_source}.

    \subsection{\label{sec:bremsstrahlung}Bremsstrahlung}
    
In \cite{Luna2016TheHoles}, the metric for an accelerating black hole was found to be the double copy of the gauge field for an accelerating point charge. Indeed, for a mass $M$ on a timelike worldline $y^\mu(\tau)$ with velocity $\lambda^\mu = \d{y^\mu}{\tau}$, the associated Kerr-Schild metric is given by \cite{Luna2016TheHoles}
\begin{align}\label{eq:bremsstrahlung_ks}
    \phi &= \frac{r_s}{r}, \qquad k^\mu(x) = \left.\frac{x^\mu - y^\mu(\tau)}{r}\right|_\text{ret}, \qquad r = -\left. \lambda^\mu(\tau) (x_\mu-y_\mu(\tau))\right|_\text{ret},
\end{align}
where $f(x,\tau)|_\text{ret}$ refers to $f(x,\tau)$ evaluated at the time $\tau$ at which a past light cone from $x$ intersects the worldline $y^\mu(\tau)$. 

Note that this reduces to the Schwarzschild solution when $y^\mu(\tau) = (\tau, 0, 0, 0)$. The constant timelike Killing field used in the Schwarzschild solution can be understood as $\left.\lambda^\mu\right|_\text{ret}$ for this particular choice of $y^\mu(\tau)$. In the accelerating case, we will continue to use $\ell^\mu \equiv \left.\lambda^\mu\right|_\text{ret}$ as a geodesic field to form the single copy. However, in this case $\ell^\mu$ is not a Killing field of the background, so we will have to use the full version (\ref{eq:sources_dbl_copy}).

Substituting the single copy gauge field $A^\mu = \frac{qg}{4\pi r}k^\mu$ into the Maxwell equations, we find a current source
\begin{equation}\label{eq:bremsstrahlung_current}
    j^\mu(x) = \left.2\frac{q}{4\pi r^2}(k\cdot \dot{\lambda})k^\mu\right|_\text{ret} - q \int d\tau\, \delta^{(4)}(x-y(\tau))\lambda^\mu.
\end{equation}

The source of the bremsstrahlung metric contains, in addition to the singularity at the mass, a trace-free term \cite{Luna2016TheHoles}:
\begin{equation}
    {T^\mu}_\nu(x) = \left.\frac{3M}{4\pi}\frac{k\cdot \dot{\lambda}(\tau)}{r^2}k^\mu k_\nu\right|_\text{ret} + M\int d\tau\,\delta^{(4)}(x-y(\tau)) \lambda^\mu \lambda_\nu,
\end{equation}
where $\dot{\lambda}^\mu = \d{\lambda^\mu}{\tau}$. Note that, since the stress-energy tensor is supported on the whole spacetime, our argument for the geodecity of $k^\mu$ from section \ref{sec:ks} breaks down. Nonetheless, since retarded quantities are constant along the integral curves of $k^\mu$, we have $k^\mu \partial_\mu k^\nu = 0$. 

We can now evaluate $\hat{\jmath}^\mu$. We will need the identities \citep{Luna2016TheHoles}
\begin{align}\label{eq:br_identities}
\begin{split}
    \partial_\mu k_\nu &= -\frac{1}{r}\left(\eta_{\mu\nu}-\lambda_\mu k_\nu - k_\mu \lambda_\nu + k_\mu k_\nu(1 + rk\cdot\dot{\lambda})\right),\\
    \partial_\mu r &= -\lambda_\mu + k_\mu(1+rk\cdot\dot{\lambda}), \\
    \partial_\mu \lambda^\nu &= k_\mu \dot{\lambda}^\nu.
\end{split}
\end{align}
Given that the background is flat, and that $\partial_{[\mu}k_{\nu]} = 0$, we can write $\hat{\jmath}^\mu$ as
\begin{equation}
    \hat{\jmath}^\mu = -\frac{q}{\kappa M(\ell\cdot k)}\lambda^\nu \big(\left(\partial^\mu(\phi k^\sigma)-\partial^\sigma(\phi k^\mu)\right)\partial_\sigma k_\nu + \partial_\sigma\left(2\phi k^\sigma \partial^\mu k_\nu + k^\mu k^\sigma \partial_\nu \phi\right)\big).
\end{equation}
Using \eqref{eq:br_identities}, we can compute the contractions
\begin{align}
    \lambda^\nu \partial_\sigma k_\nu &= -(k\cdot \dot{\lambda})k_\sigma, \qquad k^\sigma \partial_\sigma \phi = \frac{\phi}{r}, \qquad \lambda^\sigma \partial_\sigma \phi = -\phi (k\cdot\dot{\lambda}).
\end{align}
Using these, it is straightforward to compute
\begin{equation}
    \hat{\jmath}^\mu = \frac{q}{\pi r^2}(k\cdot\dot{\lambda}).
\end{equation}
It follows that the total gauge current is
\begin{equation}
    j^\mu = -\frac{2q}{M}\ell^\nu \left({T^\mu}_\nu - \frac{T}{2}{\delta^\mu}_\nu\right) + \hat{\jmath}^\mu = \frac{q(k\cdot\dot{\lambda})}{2\pi r^2}k^\mu - q\int d\tau \delta^{(4)}(x-y(\tau))\lambda^\mu,
\end{equation}
in agreement with \eqref{eq:bremsstrahlung_current}.

\section{Integrating the Kerr Source}\label{sec:kerr_ks}

With some added difficulty relative to the Schwarzschild geometry, we can use the double copy procedure to extract the Kerr-Schild form of the Kerr metric from a corresponding gauge theory source. Our starting point is the current obtained via equation (\ref{eq:sources_dbl_copy}). However, if we attempt to integrate \eqref{eq:kerr_current} to obtain a gauge field, we will find a divergent result owing to the singularity on the ring $r = 0$, $\theta = \frac{\pi}{2}$. This divergence stems from the derivation of the Kerr source in \cite{Israel1970SourceMetric}, which does not carefully account for the distributional nature of the stress-energy tensor. A more delicate treatment appears in \citep{Balasin1994DistributionalFamily}, using the Cartesian coordinates
\begin{align}
    \begin{split}
        x &= \sqrt{r^2+a^2}\sin\theta\cos\psi,\\
        y &= \sqrt{r^2+a^2}\sin\theta\sin\psi,\\
        z &= r\cos\theta.
    \end{split}
\end{align}
The Ricci tensor is computed to be
\begin{align}
\begin{split}
    {R^0}_0 &= 2M\left\lbrack \frac{a \vartheta(a-\rho)}{(a^2-\rho^2)^{3/2}}\right\rbrack\delta(z) - \frac{2M}{a}\delta(\rho-a)\delta(z),\\
    {R^0}_i &= 2M\left\lbrack \frac{\rho \vartheta(a-\rho)}{(a^2-\rho^2)^{3/2}}\right\rbrack\delta(z)(d\psi)_i - \frac{\pi M}{a}\delta(\rho-a)\delta(z) (d\psi)_i,\\
    {R^i}_j &= -\frac{2M}{a}\left\lbrack \frac{\rho^2 \vartheta(a-\rho)}{(a^2-\rho^2)^{3/2}}\right\rbrack\delta(z)(d\psi)_i(d\psi)_j + \frac{2M}{a}\frac{\vartheta(a-\rho)}{(a^2-\rho^2)^{1/2}}\delta(z) e^i_z e^j_z \\ &+ \frac{4M}{a}\delta(\rho-a)\delta(z)(d\psi)_i(d\psi)_j
\end{split}
\end{align}
where the distribution $\left\lbrack \frac{\vartheta(a-\rho)}{(a^2-\rho^2)^{3/2}}\right\rbrack$ is defined by
\begin{equation}
    \left(\left\lbrack \frac{\vartheta(a-\rho)}{(a^2-\rho^2)^{3/2}}\right\rbrack,\varphi\right) = \int_0^{2\pi}d\theta\,\int_0^a \rho\,d\rho\,\frac{1}{(a^2-\rho^2)^{3/2}}(\varphi(\rho)-\varphi(a)).
\end{equation}
Since $\hat{\jmath}^\mu = 0$, the current is
\begin{equation}
    j^\mu = \frac{2q}{\kappa M}R^\mu_0.
\end{equation}
We will start by determining $A^0$. Since the Kerr metric is stationary, gauge transformations will not change $A^0$ except by addition of a constant, and this constant is fixed by the boundary condition $\phi(r)\to 0$ as $r\to \infty$. Thus, we have
\begin{equation}
    A^0(x) = -g\int d^3 x'\,\frac{j^0(x')}{4\pi|x-x'|}.
\end{equation}
The distance between a point $x$ with spheroidal coordinates $(r\equiv \lambda a, \theta)$ and a point $x'$ on the disk with polar coordinates $(\rho\equiv sa, \psi)$ is
\begin{equation}
    |x-x'| = a\sqrt{s^2+\lambda^2 + \sin^2\theta -2s\sqrt{\lambda^2+1}\sin\theta\cos\psi}.
\end{equation}
Therefore,
\begin{equation}\label{eq:kerr_a0_integral}
    A^0(x) = \frac{gq}{8\pi^2 a}\left(\int_0^{2\pi}d\psi\,\int_0^1 \,ds\,\frac{s}{(1-s^2)^{3/2}}\left(f(s,\psi)-f(1,\psi)\right) - \int_0^{2\pi}d\psi\,f(1,\psi)\right),
\end{equation}
where
\begin{equation}
    f(s,\psi) = \left(s^2 -2s\sqrt{\lambda^2+1}\sin\theta\cos\psi + \lambda^2+\sin^2\theta\right)^{-1/2}.
\end{equation}
On the symmetry axis, where $\theta = 0$, the integral is simple enough to evaluate analytically. We find
\begin{equation}
    A^0(r,\theta = 0) = \frac{gq}{4\pi a}\frac{\lambda}{1+\lambda^2} = \frac{gq}{4\pi}\frac{r}{r^2+a^2}.
\end{equation}
Off axis, we evaluate the integral numerically, and find that it continues to agree with the relation $A^0 = \frac{gq}{\kappa M}\phi k^0$:
\begin{equation}
    A^0(r, \theta) = \frac{gq}{4\pi}\frac{r}{r^2+a^2\cos^2\theta}.
\end{equation}

It is straightforward to show that in Lorenz gauge, the only other nonzero component of the gauge field will be $A^\psi$. We can obtain this component from an integral similar to \eqref{eq:kerr_a0_integral}, with an additional factor of $\frac{\cos\psi}{\sqrt{1+\lambda^2}\sin\theta}$ to account for the direction of the unit $\phi$ vector on the disk, and the scale factor in the $\psi$ direction at the point $x$. The integral is
\begin{align}
\begin{split}
    A^\psi(x) = \frac{gq}{8\pi^2 a}\frac{1}{\sqrt{1+\lambda^2}\sin\theta}\Bigg(&\int_0^{2\pi}d\psi\,\int_0^1 ds\,\frac{s^2\cos\psi}{(1-s^2)^{3/2}}(f(s,\psi)-f(1,\psi))\\
    -\frac{\pi}{2}&\int_0^{2\pi}d\psi\,f(1,\psi)\cos\psi\Bigg).
\end{split}
\end{align}
Numerically, we find that this gives the result
\begin{equation}
    A^\psi(r,\theta) = -\frac{gq}{4\pi}\frac{r}{r^2+a^2\cos^2\theta}\frac{a}{r^2+a^2}
\end{equation}
which we expect from the form of the Kerr-Schild transformation in section \ref{sec:kerr}.

In order to form a Kerr-Schild transformation from the gauge vector we have found, we need to find a gauge transformation which makes $(A^\mu+\partial^\mu\chi)(A_\mu+\partial_\mu\chi) = 0$. Since only $A^0$ and $A^\psi$ are nonzero, we can write $A^\mu$ as a linear combination of the Killing vectors $(\partial_t)^\mu$ and $(\partial_\psi)^\mu$, and it follows that $A^\mu\partial_\mu\chi = 0$. Therefore, we need to find a function $\chi(r, \theta)$ such that
\begin{equation}
    (\partial_i\chi)(\partial^i\chi) = -A^\mu A_\mu = \left(\frac{gq}{4\pi}\right)^2\frac{r^2}{(r^2+a^2)(r^2+a^2\cos^2\theta)}.
\end{equation}
Since $g^{rr} = \frac{r^2+a^2}{r^2+a^2\cos^2\theta}$, we see that we should have $\partial_\theta\chi = 0$ and $\partial_r\chi = \frac{gq}{4\pi}\frac{r}{r^2+a^2}$, so 
\begin{equation}
    \chi = \frac{gq}{8\pi}\log\left(\frac{r^2+a^2}{r_s^2}\right).
\end{equation}
After the gauge transformation, we have
\begin{equation}
    A^\mu = \frac{gq}{4\pi}\frac{r}{r^2+a^2\cos^2\theta}\left(1, 1, 0, -\frac{a}{r^2+a^2}\right).
\end{equation}
We have assumed throughout that the single copy is formed with the unit timelike Killing vector $\ell^\mu$ and that $\ell\cdot k = -1$, so we have
\begin{align}
    \phi &= \frac{\kappa M}{qg}A^0 = \frac{r_s r}{r^2+a^2\cos^2\theta}, \qquad k^\mu = \left(1, 1, 0, -\frac{a}{r^2+a^2}\right),
\end{align}
exactly as in \eqref{eq:kerr_ks}.

\end{appendices}

\bibliographystyle{JHEP}
\bibliography{main}

\end{document}